\documentclass[preprint,10pt]{elsarticle}
\usepackage{amsfonts}
\usepackage{epsfig,amsmath,latexsym,amssymb}
\usepackage{amscd}
\usepackage{multirow}
\usepackage{graphicx}
\usepackage{geometry}
\usepackage{lscape}
\usepackage{amsmath}
\usepackage{amssymb}
\usepackage{bm}
\usepackage{subfigure}
\usepackage{url}
\usepackage{pifont}
\usepackage{bbding}
\usepackage{color}
\usepackage{natbib}
\usepackage{float}
\usepackage{placeins}
\usepackage{bm}

\providecommand{\U}[1]{\protect\rule{.1in}{.1in}}
\def\E{{\mathbb E}}

\newcommand{\Remm}[1]{}
\usepackage{placeins}
\newtheorem{theo}{Theorem}[section]
\newtheorem{prop}[theo]{Proposition}
\newtheorem{lem}[theo]{Lemma}
\newtheorem{cor}[theo]{Corollary}
\newtheorem{exam}[theo]{Example}

\newtheorem{rem}[theo]{Remark}
\newtheorem{defi}[theo]{Definition}

\usepackage{dsfont}
\usepackage{url}

\numberwithin{equation}{section}
\renewcommand{\thesection}{\arabic{section}}

\begin{document}

\begin{frontmatter}

\title{Understanding Operational Risk Capital Approximations: First and Second Orders}
\author{Gareth W.~Peters$^{1,2,3}$ \quad Rodrigo S.~Targino$^{1}$ \quad Pavel V.~Shevchenko$^{2}$} 
\date{{\footnotesize {Working paper, version from \today }}}
\maketitle

%%%%%%%%%%%%%%%%%%%%%%%%%%%%%%%%%%%%%%%%%%%%%%%%%%%%%%%%%%%%%%%%%%%%%%%%%%%%%%%%%%%%%%%%%%
%%%%%%%%%%%%%%%%%%%%%%%%%%%%%%%%%%%%%%%%%%%%%%%%%%%%%%%%%%%%%%%%%%%%%%%%%%%%%%%%%%%%%%%%%%
\begin{abstract}
%%%%%%%%%%%%%%%%%%%%%%%%%%%%%%%%%%%%%%%%%%%%%%%%%%%%%%%%%%%%%%%%%%%%%%%%%%%%%%%%%%%%%%%%%%
%%%%%%%%%%%%%%%%%%%%%%%%%%%%%%%%%%%%%%%%%%%%%%%%%%%%%%%%%%%%%%%%%%%%%%%%%%%%%%%%%%%%%%%%%%

\noindent We set the context for capital approximation within the framework of the Basel II / III regulatory capital accords. This is particularly topical as the Basel III accord is shortly due to take effect. In this regard, we provide a summary of the role of capital adequacy in the new accord, highlighting along the way the significant loss events that have been attributed to the Operational Risk class that was introduced in the Basel II and III accords. Then we provide a semi-tutorial discussion on the modelling aspects of capital estimation under a Loss Distributional Approach (LDA). Our emphasis is to focuss on the important loss processes with regard to those that contribute most to capital, the so called ``high consequence, low frequency'' loss processes. 

This leads us to provide a tutorial overview of heavy tailed loss process modelling in OpRisk under Basel III, with discussion on the implications of such tail assumptions for the severity model in an LDA structure. This provides practitioners with a clear understanding of the features that they may wish to consider when developing OpRisk severity models in practice. From this discussion on heavy tailed severity models, we then develop an understanding of the impact such models have on the right tail asymptotics of the compound loss process and we provide detailed presentation of what are known as first and second order tail approximations for the resulting heavy tailed loss process. From this we develop a tutorial on three key families of risk measures and their equivalent second order asymptotic approximations: Value-at-Risk (Basel III industry standard); Expected Shortfall (ES) and the Spectral Risk Measure. These then form the capital approximations.

We then provide a few example case studies to illustrate the accuracy of these asymptotic captial approximations, the rate of the convergence of the assymptotic result as a function of the LDA frequency and severity model parameters, the sensitivity of the capital approximation to the model parameters and the sensitivity to model miss-specification.

\vspace{5mm}
\end{abstract}


\begin{center}
{\footnotesize {\ \textit{$^{1}$Department of Statistical Science, University College London; \\[0pt]
email: gareth.peters@ucl.ac.uk \\[0pt]
(Corresponding Author) \\[0pt]
$^{2}$ CSIRO Mathematics, Informatics and Statistics, Sydney, Australia \\[0pt]
$^{3}$ Associate Fellow, Oxford-Man Institute, Oxford University \\[0pt] } } }
\end{center}

\end{frontmatter}

%%%%%%%%%%%%%%%%%%%%%%%%%%%%%%%%%%%%%%%%%%%%%%%%%%%%%%%%%%%%%%%%%%%%%%%%%%%%%%%%%%%%%%%%%%%%%%%%%%
%%%%%%%%%%%%%%%%%%%%%%%%%%%%%%%%%%%%%%%%%%%%%%%%%%%%%%%%%%%%%%%%%%%%%%%%%%%%%%%%%%%%%%%%%%%%%%%%%%
\section{The Changing Landscape of Capital Accords}
%%%%%%%%%%%%%%%%%%%%%%%%%%%%%%%%%%%%%%%%%%%%%%%%%%%%%%%%%%%%%%%%%%%%%%%%%%%%%%%%%%%%%%%%%%%%%%%%%%
%%%%%%%%%%%%%%%%%%%%%%%%%%%%%%%%%%%%%%%%%%%%%%%%%%%%%%%%%%%%%%%%%%%%%%%%%%%%%%%%%%%%%%%%%%%%%%%%%%

\begin{figure}
\includegraphics[width=\textwidth, height=10cm]{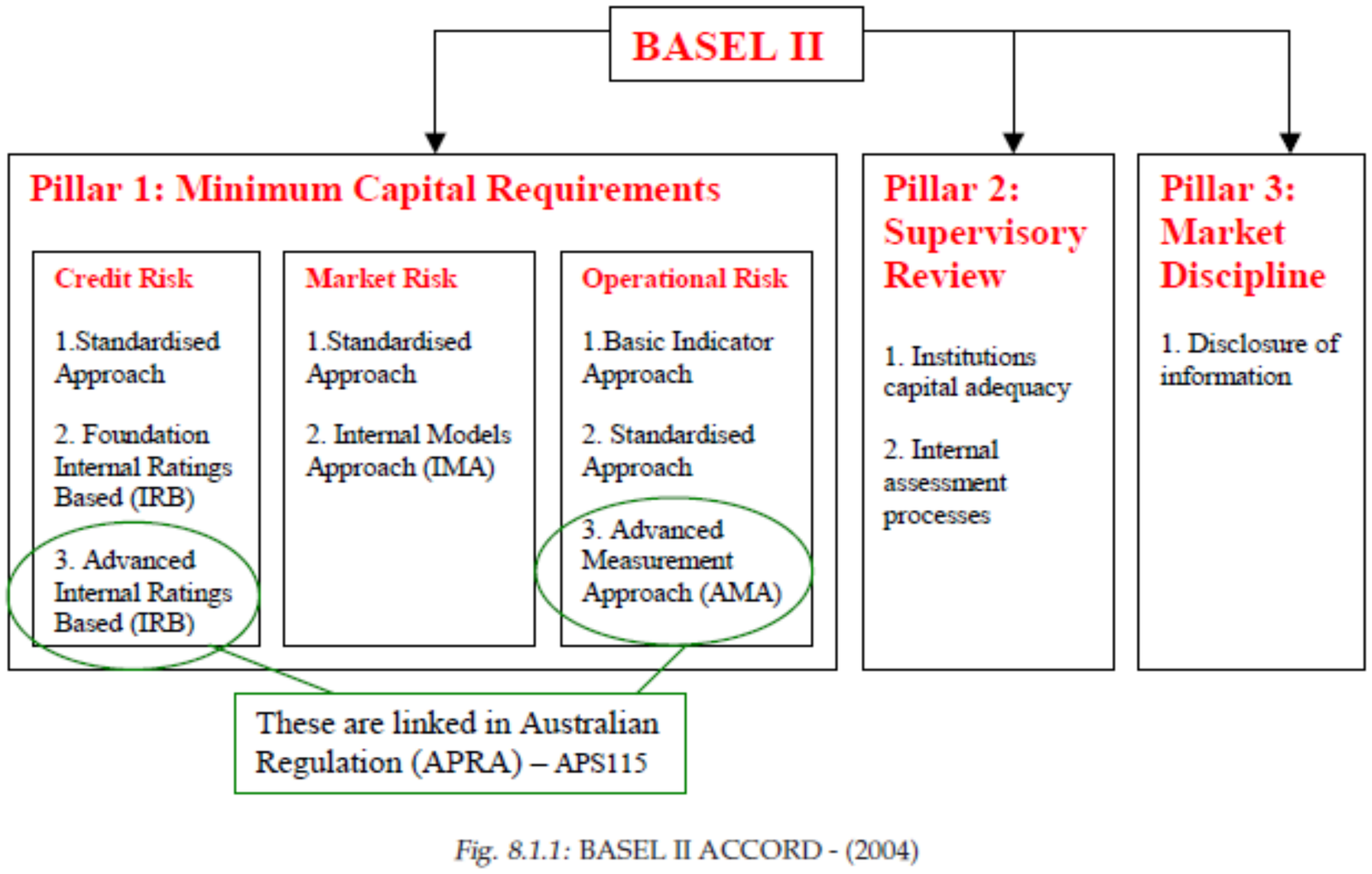}
\caption{The Basel II and now Basel III pillars for capital adequacy assessment.}
\label{fig1}
\end{figure}

In juristictions in which active regulation is applied to the banking sector throughout the world, the modelling of Operational Risk (OpRisk) has progressively taken a prominent place in financial quantitative measurement. This has occurred as a result of Basel II and now Basel III regulatory requirements. For example in the context of banking regulation in Australia, the basic framework of Basel II/III is summarized in Figure \ref{fig1}. In this juristiction one observes that a large amount of the developments in quantitative methodology for estimation of OpRisk capital, development of OpRisk frameworks embedded within retail banks and large financial institutions as well as infrastructure for collection and reporting of losses in data bases has been achieved largely due to regulatory choices to link Advanced Measurement Approaches in OpRisk modelling to Credit Risk. In other areas the progression of such features has lagged behind the Australian example but now OpRisk in other large banking sectors is becomming increasingly prominent.

There has been a significant amount of research dedicated to understanding the features of Basel II, see for example \cite{danielsson2001academic}, \cite{decamps2004three} and \cite{kashyap2004cyclical}. In addition the mathematical and statistical properties of the key risk processes that comprise OpRisk, especially those that contribute significantly to the capital charge required to be held against OpRisk losses have also been carefully studied, see for example the book length discussions in \cite{cruz2002modeling}, \cite{king2001operational} and \cite{shevchenko2011modelling}. 

There have been both numerical and simulation based approaches adopted as well as analytical mathematical developments of closed form approximations for capital approximation. In this manuscript we aim to provide a clear and concise understanding of several increasingly popular approaches to capital approximation from an analytic perspective. In the process we demonstrate several important key details of such capital approximations, we discuss the implications of use of such approximations, their shortcommings and we assess their behaviours in simple but realistic loss models adopted in practice. 

In particular we provide a detailed discussion on the attributes of the so called capital approximation methods known as Single Loss type approaches of first and second order expansions. In doing so, this manuscript draws together several disperate areas of the literature to allow industry professionals an insight into the processes that have lead to expansions of tail functionals for risk measures such as Value-at-Risk, Expected-Shortfall and Spectral Risk measures, considered for capital definitions.

In understanding the context for these developments we discuss first the regulatory evolution of OpRisk modelling frameworks. In January 2001 the Basel Committee on Banking Supervision proposed the Basel II Accord (\cite{BCBS1},\cite{BCBS2}, \cite{BCBS3}) which replaced the 1988 Capital Accord. Now in 2013 the Basel III Accord is due to start to be considered. Since the initiation of the Basel captial accords, the discipline of OpRisk and its quantification have grown in prominence in the financial sector. Paralleling these developments have been similar regulatory requirements for the insurance industry which are referred to as Solvency 2. 

Under the Basel II/III structures there is at the core the notion of three pillars, which, by their very nature, emphasize the importance of assessing, modelling and understanding OpRisk loss profiles. These 3 pillars are; minimum capital requirements (refining and enhancing riskmodelling frameworks), supervisory review of an institution’s capital adequacy and internal assessment processes and market discipline, which deals with disclosure of information, see Figure \ref{fig1}. 

In the third update to the Basel Accords due for implementation in the period 2013-2018, a global regulatory standard which draws together bank capital adequacy, stress testing and market liquidity is created. It is established as an international best practice for modelling OpRisk by the members of the Basel Committee on Banking Supervision, see \cite{BCBS4} and discussions in \cite{blundell2010thinking}.

The Basel III accord naturally extends on the work developed in both the Basel I and Basel II accords with the new accord arising primarily as a response to the identified issues associated with financial regulation that arose during the recent global financial crisis in the late-2000s. In this regard, the Basel III accord builds on Basel II by strengthening the bank capital requirements as well as introducing additional regulatory requirements on bank liquidity and leverage.

Banking regulation under Basel II and Basel III specifies that banks are required to hold adequate capital against OpRisk losses. OpRisk is a relatively new category of risk which is additional to more well established risk areas such as market and credit risks. As such in its own right OpRisk attracts a capital charge which is defined by Basel II/III [1, p.144] as: ``\textit{[. . . ] the risk of loss resulting from inadequate or
failed internal processes, people and systems or from external events. This definition includes
legal risk, but excludes strategic and reputational risk.}''
OpRisk is significant in many financial institutions. 

Before detailing the changes to capital requirements due to come into industry practice under Basel III, it is prudent to recall the Basel definition of Tier I capital, which is the key measure of a bank's financial strength from the perspective of the regulatory authority. In particular the capital accord in Basell II and III states that financial institutions must provide capital above the minimum required amount, known as the floor capital. In addition this capital as specified in regulation is comprised of three key components, Tier I, Tier II and Tier III. Both Tier I and Tier II capital were first defined in the Basel I capital accord and remained substantially the same in the replacement Basel II and Basel III accords.

\begin{defi}[Tier I Capital] The Tier I capital under regulation is comprised of the following main components:
\begin{enumerate}
\item Paid-up share capital / common stock;
\item Disclosed Reserves (or retained earnings).
\end{enumerate}
It may also include non-redeemable non-cumulative preferred stock. 
\end{defi}

The Basel Committee also noted the existance of banking strategies to develop instruments in order to generate Tier I capital. As a consequence, these must be carefully regulated through imposition of stringent conditions, with a limit to such instruments at a maximum of 15\% of total Tier I capital. 

\begin{defi}[Tier II Capital] The Tier II capital under regulation is comprised of the following main components:
\begin{enumerate}
\item Undisclosed reserves;
\item Asset revaluation reserves;
\item General provisions / general loan-loss reserves;
\item Hybrid (debt/equity) capital instruments; and
\item Long-term subordinated debt.
\end{enumerate}
In this regard one may consider Tier II capital as representing so called "supplementary capital". 
\end{defi}
 
We note at this stage that as a consequence of different legal systems in each juristiction, the accord has had to be sufficiently flexible to allow for some interpretation of specific capital componets within the context of each regulators juristiction. Depending on the particular juristiction in question, the specific country's banking regulator has some discretionary control over how exactly differing financial instruments may count in a capital calculations. 

\begin{rem} The key reason that Basel III requires financial institutions to hold capital is that it is aimed to provide protection against unexpected losses. This is different to mitigation of expected losses, which are covered by provisions, reserves and current year profits. 
\end{rem}

We note that modifications under the Basel III accord relative to its predecessor refer to limitations on Risk Weighted Capital (RWC) and the Tier I Capital Ratio, as defined below.

\begin{defi}[Risk Weighted Assets] These assets comprise the total of all assets held by the bank weighted by credit risk according to a formula determined by either the juristictions regulatory authority or in some cases the central bank. Most regulators and central banks adhere to the definitions specified by the Basel Committee on Banking Supervision (BCBS) guidelines in setting formulae for asset risk weights. Liquid assets such as cash and coins typically have zero risk weight, while certain loans have a risk weight at 100\% of their face value. As specified by the BCBS the total RWA is not limited to Credit Risk. It contains components for Market Risk (typically based on value at risk (VAR) ) and Operational Risk. The BCBS rules for calculation of the components of total RWA have also been updated as a result of the recent financial crisis.
\end{defi}

\begin{defi}[Tier I Capital Ratio]
The Tier 1 capital ratio is the ratio of a bank's core equity capital to its total risk-weighted assets (RWA). 
\end{defi}

Next we highlight the prominent extensions to the Basel II accord, established in the Basel III accord. In particular the Basel III accord will require financial institutions to hold for risk weighted assets, 4.5\% of common equity which is an increase from the previous 2\% under Basel II as well as 6\% of Tier I capital itself an increase by 2\% relative to Basel II. In addition to these changes to common equity and Tier I capital, Basel III also introduces a minimum leverage ratio and two additional required liquidity ratio limits. Finally, of the significant changes, there are also additional capital buffers introduced:
\begin{enumerate} 
\item Introduction of a mandatory capital conservation buffer of 2.5\% ; and 
\item A discretionary countercyclical buffer, allowing national regulators to require up to another 2.5\% of capital during periods of high credit growth. 
\end{enumerate}
%REFERENCE ^ http://www.bis.org/publ/bcbs189.pdf

%%%%%%%%%%%%%%%%%%%%%%%%%%%%%%%%%%%%%%%%%%%%%%%%%%%%%%%%%%%%%%%%%%%%%%%%%%%%%%%%%%%%%%%%%%%%%%%%%%
\subsection{Understanding the Significance of OpRisk Losses}
%%%%%%%%%%%%%%%%%%%%%%%%%%%%%%%%%%%%%%%%%%%%%%%%%%%%%%%%%%%%%%%%%%%%%%%%%%%%%%%%%%%%%%%%%%%%%%%%%%
To illustrate the significance of OpRisk losses on the stability of banking operations and the extent that particular loss processes in this class of risk can threaten financial insolvency for banks we note some aggregate loss figures and illustrate how such aggregates are possible by noting particular instances of prominent loss events that have occurred under the OpRisk class. 

It was reported in \cite{gagan2008operational} that the total loss associated with operational risk has reached as high as US\$ 96 billion in the United States during the financial crisis in 2008. There have also been numerous OpRisk loss events that have been highlighted in the media to support such enormous aggregate figures. Such single event examples of extremely large OpRisk losses include: Barings Bank (loss GBP 1.3 billion in 1995), Sumitomo Corporation (loss USD 2.6 billion in 1996), Enron (USD 2.2 billion in 2001), and recent loss in Society Generale (Euro 4.9 billion in 2008).  

Some of the lesser reported cases have recently come to light with the paper of \cite{LuGuo2012} who paint similar pictures in the Chinese banking sectors as have been observed in US and European markets. For example they state that typical examples of large OpRisk loss events in recent years in the Chinese banking sector include the Guangdong branch of the Industrial and Commercial Bank of China (ICBC) which in 2003 lost 740 million yuan; the Jinzhou branch of the Bank of Communications in 2004 which lost 22.1 million yuan; the Heilongjiang branch of the Bank of China (BOC) in 2005 which lost 100 million yuan; the Guangdong branch of BOC in 2006  lost 400 million yuan; and the Qilu Bank in 2010 which lost 100 million yuan. 

Each of these single loss events are significant and indicate the importance of models for loss processes which will capture such extreme loss events adequately when undertaking capital estimation. These illustrative examples and many more, has provided a clear focus for practitioners and risk modellers to invest in efforts to understand heavy tailed loss modelling, which has been highlighted in numerous reviews on OpRisk modelling, see \cite{moscadelli2004modelling}, \cite{nevslehova2006infinite}, \cite{peters2011analytic}, \cite{Peters2006}, \cite{giacometti55heavy} and \cite{dutta2006tale}. In particular we will focus on the most widely used model framework involving a compound process reprsentation of the risk process. 

\FloatBarrier

%%%%%%%%%%%%%%%%%%%%%%%%%%%%%%%%%%%%%%%%%%%%%%%%%%%%%%%%%%%%%%%%%%%%%%%%%%%%%%%%%%%%%%%%%%%%%%%%%%
\subsection{ A Brief Background on Loss Distributional Approach (LDA) Models}
%%%%%%%%%%%%%%%%%%%%%%%%%%%%%%%%%%%%%%%%%%%%%%%%%%%%%%%%%%%%%%%%%%%%%%%%%%%%%%%%%%%%%%%%%%%%%%%%%%

To quantify the operational risk capital charge under the current regulatory framework for banking supervision, referred to as Basel II/ Basel III, many banks adopt the Loss Distribution Approach (LDA). There are several modeling issues that should be resolved to use this approach in practice, a detailed review on the quantitative properties of estimation can be found in \cite{dutta2006tale}, \cite{shevchenko2009implementing}, \cite{peters2007simulation}, \cite{peters2006bayesian}, \cite{peters2009dynamic}, \cite{peters2009dynamic}, \cite{chernobai2004note} and in addition in some important heavy tailed settings (large consequence, rare occurance) closed form representations of such models in \cite{peters2011impact}, \cite{peters2011analytic} and \cite{nevslehova2006infinite}.

In this section we fist motivate and introduce the context of LDA modeling in risk and insurance. Next, we provide a brief specifically selected survey of closed form analytic results known in the actuarial and risk literature for sub-classes of such LDA models as the Single Loss Approximations (hereafter SLA). As pointed out previously it is precisely these heavy tailed loss processes that result in the significant individual loss events, as discussed above. We therefore then focus on key elements of heavy tailed loss process asymptotics with a view to understanding capital approximations.
In doing so we draw together several disparate sources of information for practitioners from sources in both mathematics and risk literature, for example we consider results recently developed in actuarial literature for the heavy tailed case corresponding to the first order and second order asymptotic approximations, see comprehensive discussions in a general context in \cite{albrecher2010higher}, \cite{degen2010calculation} and the books, \cite{barbe2009asymptotic} and the forthcoming \cite{CruzPetersShevchenko}. 

We conclude this section by observing that according to regulatory standards and indeed good risk management practice \textit{such asymptotic SLA approximations are often required to be accompanied with numerical and statistical solutions which can more readily take into account \underline{model uncertainty}, \underline{parameter uncertainty}, \underline{parameter sensitivity} and \underline{asymptotic rate of convergence analysis}}, see discussions in \cite{del2012introduction}.

 The fact that such approximations are inherently asymptotic in nature, and may be inaccurate outside of the neighborhood of infinity, means such analysis is directly relevant in practice when considering the suitability of such approximations for capital calculations. It is important that these be both accurate and relatively stable over time. However, incorporating these features into a SLA is often highly challenging. 

To begin, consider the widely utilized insurance model known as a single risk LDA model. This represents the standard under the Basel II/III capital accords \cite{basel2001basel} and involves an annual loss in a risk cell (business line/event type) modeled as a compound distributed random variable,
\begin{equation}
Z_{t}^{\left( j\right) }=\sum\limits_{s=1}^{N_{t}^{\left( j\right)
}}X_{s}^{\left( j\right) }\left( t\right),  
\label{AnnLoss1}
\end{equation}
for $t=1,2,\ldots,T$ discrete time (in annual units) and index $j$ identifies the risk cell. Furthermore, the annual number of losses is denoted by $N_{t}^{(j)}$ which is a random variable distributed according to a frequency counting distribution $P^{(j)}(\cdot) $, typically Poisson, Binomial or Negative Binomial. The severities (losses) in year $t$ are represented by random variables $X_{s}^{(j)}(t)$, $s \ge 1$, distributed according to a severity distribution $F^{(j)}(\cdot)$ and there are $N^{(j)}_t$ of them in year $t$.

Before proceeding, it will be relevant to define some basic notation adopted throughout. In general, we will suppress the risk cell index $j$ and time index $t$ unless explicitly utilised. Therefore, we denote by $F(x)$ a distribution with positive support for the severity model characterizing the loss distribution for each random variable $X_{s}$ for $s \in \left\{1,2,\ldots,N\right\}$. We denote the annual loss (aggregated loss) by $Z$ with annual loss distribution $G = F_Z$ and the partial sum of $n$ random losses by $S_n$ with distribution $F_{S_n}=F^{n*}$ where $F^{n*}$ denotes the $n$-fold convolution of the severity distribution for the independent losses. The densities, when they exist, for the severity distribution and annual loss distributions will be denoted by $f(x)$ and $f_Z(x)$ respectively. 

In constructing the LDA model we assume that all losses are i.i.d. with $X_{s}^{\left( j\right) }\left( t\right) \sim F(x)$ and that the severity distribution is continuous with no atoms in the support $[0,\infty)$. As a consequence, linear combinations (aggregation) of losses in a given year, denoted by the partial sum
$$
S_n(t) = \sum_{s=1}^n X_{s}^{\left( j\right) }\left( t\right) \sim F_{S_n}(x)
$$ 
have the following analytic representation:
\begin{equation*}
\begin{split}
F_{S_n}(x) &= \left(F \star F \star \cdots F\right)(x) = \int_{[0,\infty)}F^{(n-1)\star}(x-y)dF(x).
\end{split}
\end{equation*}
In \cite{feller1966introduction} it is shown that if $F(x)$ has no atoms in $[0,\infty)$ then the $n$-fold convolution of such severity distributions will also admit no atoms in the support $[0,\infty)$. In addition we note that continuity and boundedness of a severity distribution $F(x)$ is preserved under $n$-fold convolution. Hence, if $F(x)$ admits a density $\frac{d}{dx}F(x)$ then so does the distribution of the partial sum $F_{S_n}$, for any $n \in \left\{1,2,\ldots\right\}$ and compound process (random sum) $F_{Z}$. For most models such analytic representations of the combined loss distribution are non closed form, with the exception of special sub-families of infinitely divisible severity distribution models, see \cite{peters2011analytic}.

%%%%%%%%%%%%%%%%%%%%%%%%%%%%%%%%%%%%%%%%%%%%%%%%%%%%%%%%%%%%%%%%%%%%%%%%%%%%%%%%%%%%%%%%%%%%%%%%%%
%%%%%%%%%%%%%%%%%%%%%%%%%%%%%%%%%%%%%%%%%%%%%%%%%%%%%%%%%%%%%%%%%%%%%%%%%%%%%%%%%%%%%%%%%%%%%%%%%%
\section{ On the Road to Capital Approximation: a Tale of Tails}
%%%%%%%%%%%%%%%%%%%%%%%%%%%%%%%%%%%%%%%%%%%%%%%%%%%%%%%%%%%%%%%%%%%%%%%%%%%%%%%%%%%%%%%%%%%%%%%%%%
%%%%%%%%%%%%%%%%%%%%%%%%%%%%%%%%%%%%%%%%%%%%%%%%%%%%%%%%%%%%%%%%%%%%%%%%%%%%%%%%%%%%%%%%%%%%%%%%%%

In this section we present an overview of important technical results from the probability and mathematical statistics literature that will lead to an understanding of OpRisk capital approximation techniques that are being discussed in the OpRisk literature recently, see \cite{bocker2005operational}, \cite{bocker2006operational}, \cite{bocker2009first}, \cite{hess2011can} and \cite{degen2010calculation}. Since this section is aimed at a guided review we make explicit some important definitions that are used throughout, for example we remind the reader of the notion of asymptotic equivalence and max-sum equivalence, as well as several key definitions relating to tail asymptotics for a severity model under an LDA framework that are not widely known by practitioners utilising such approximations. 

\begin{defi}[Asymptotic Equivalence]
A probability distribution function $F(x)$ is \textit{asymptotically equivalent} to another probability distribution function $G(x)$, denoted by $F(x) \sim G(x)$ as $x \rightarrow \infty$ if it holds that, $\forall \epsilon > 0, \exists x_0$ such that $\forall x > x_0$ the following is true
\begin{equation}
\left|\frac{F(x)}{G(x)} - 1\right| < \epsilon.
\end{equation}
\end{defi}

\begin{defi}[Max-Sum Equivalence]
A probability distribution function is \textit{max-sum-equivalent}, denoted by $F \sim_M G$, when the convolution of the tail distribution of two random variables is distributed according to the sum of the two tail distributions asymptotically, 
$$
1 - (F \star G)(x) = (\overline{F \star G})(x) \sim \overline{F}(x) + \overline{G}(x), \;\; x \rightarrow \infty,
$$
see discussion in \cite{li2010note}.
\end{defi}

From these basic definitions, we next proceed to consider some key classifications of heavy tailed distributions. Though technical in nature, these will allow us to carefully understand the behaviour of both heavy tailed severity distributions and the compound processes constructed with these models under an LDA framework.

%%%%%%%%%%%%%%%%%%%%%%%%%%%%%%%%%%%%%%%%%%%%%%%%%%%%%%%%%%%%%%%%%%%%%%%%%%%%%%%%%%%%%%%%%%%%%%%%%%
\subsection{Review of Classifications for Heavy Tailed Severity Distributions}
%%%%%%%%%%%%%%%%%%%%%%%%%%%%%%%%%%%%%%%%%%%%%%%%%%%%%%%%%%%%%%%%%%%%%%%%%%%%%%%%%%%%%%%%%%%%%%%%%%

In practice the choice of severity distribution $F(x)$ should be considered carefully for each individual risk process as it can have a significant impact on the capital and the choice of appropriate captial approximation method. This is especially the case for those risk processes for which business managers believe there will be infrequent losses with very high consequence. It is therefore important to carefully consider the possible implications on capital calculation and capital approximation that arise when making particular assumptions about the severity distributions right tails. In this regard we begin with a basic coverage of the key features one may consider when deciding on a suitable heavy tailed severity distribution, with respect to the behaviour of the right tail, associated with probabilities of large loss amounts. We note that the classifications of severity models into particular families of heavy-taile models as discussed below is by no means supposed to be disjunctive in the understanding of such groupings. In addition we also point out that fact that we have several choices to consider when developing such high-consequence severity model assumptions. For instance we could assume properties of the right tail of the distribution function, the density function or the survival function.

For large consequence events, one may often consider distributions for which the moment generating function doesn't exist on the positive real line such that
\begin{equation}
\int e^{tx}dF(x) = \infty, \; \forall t > 0.
\end{equation}
In otherwords, the standard Markov Inequality for the expontially decaying ``light'' tail behaviour of a loss distribution in which
\begin{equation}
\overline{F}(x) \leq \exp(-sx)\mathbb{E}\left[\exp(sx)\right], \;\; \forall x > 0
\end{equation}
does not apply.

It can be shown that a distribution $F$ has a moment generating function in some right neighbourhood of the origin if and only if the following bound holds for some positive real numbers $M$ and $t$,
\begin{equation}
\overline{F}(x) \leq M\exp(-tx), \; \forall x > 0.
\end{equation}
Hence, one basic definition of an important class of distributions in OpRisk is the heavy tailed distributions which have a right tail heavier than any exponential distribution. However, there are numerous more refined categorisations of heavy tailed distributions which are required for the results in this manuscript. 

A popular class of heavy-tailed models is the sub-exponential family of severity distributions that we denote by membership $(F(x) \in \mathcal{F})$ and define below, see discussion in for example \cite{resnick2006heavy} and \cite{kluppelberg1989subexponential}.

\begin{defi}[Sub-exponential Severity Models]
The sub-exponential family of distributions $\mathcal{F}$ defines a class of heavy tailed severity models that satisfy the limits
\begin{equation}
\lim_{x \rightarrow \infty} \frac{1-F^{n\star}(x)}{1-F(x)}=n,
\end{equation}
if and only if,
\begin{equation}
\lim_{x \rightarrow \infty} \frac{1-F^{2\star}(x)}{1-F(x)}=2.
\end{equation}
\end{defi}

In \cite{pitman1980subexponential} it was demonstrated that the necessary and sufficient condition for membership of a severity distribution in the sub-exponential class $(F \in \mathcal{F})$ is satisfied if and only if the tail distribution $\overline{F}(x) = 1-F(x)$ satisfies 
$$
\lim_{x \rightarrow \infty} \int^{x}_0 \frac{\overline{F}(x-y)}{\overline{F}(x)}F(y)dy=1.
$$
Alternatively, one may characterize the family of distributions $F \in \mathcal{F}$ by those that satisfy asymptotically the tail ratio
\begin{equation}
\lim_{x \rightarrow \infty} \frac{\overline{F}(x-y)}{\overline{F}(x)} = 1, \; \forall y \in [0,\infty).
\end{equation}
Severity models $F \in \mathcal{F}$ are of interest for severity distributions in high consequence loss modeling since they include models with \textit{infinite mean loss} and \textit{infinite variance}. In addition, the class $\mathcal{F}$ includes all severity models in which the tail distribution under the log transformed r.v., $\overline{F}\left(\log(x)\right)$, is a slowly varying function of $x$ at infinity. We will discuss both regular and slow variation below.

Examples of models in this family include:
\begin{enumerate}
\item{Pareto: $\overline{F}(x) = \left(\frac{c}{\left(c+x\right)}\right)^{\alpha}$ for $x \geq 0$, $c > 0$ and $\alpha>0$}
\item{Log Normal: $\overline{F}(x) = \frac{1}{2} + \frac{1}{2}\text{erf}\left[\frac{\ln x - \mu}{\sqrt{2 \sigma^2}}\right]$, $x \geq 0$, $\mu \in \mathbb{R}$ and $\sigma > 0$.}
\item{Heavy-Tailed Weibull:  $\overline{F}(x) = \exp\left(-\lambda x^{\alpha}\right)$, $x \geq 0$, $\lambda > 0$ and $0 < \alpha < 1$.}
\end{enumerate}

We may also consider classes of heavy-tailed severity distributions as classified by their right tail properties through formal definitions such as regularly varying tail, long-tailed, dominantly varying tail, subversively varying tail and  smoothly-varying tail, each of which we briefly define below. We then relate these different classes of severity model assumptions to each other to provide a basic understanding of the relationships between each of these possible heavy tailed severity modelling assumption.

Arguably one of the most utilised sub-classes of the sub-exponential distributions is the class of regularly varying distributions. Now recalling the definition of the class of regularly varying functions given by Definition \ref{Def:RegularVaryingFn}, see \cite{Bingham1989} and \cite{resnick2006heavy}. 

\begin{defi}[Regular Variation]{\label{Def:RegularVaryingFn} A measurable function $f(x) > 0$ that satisfies the condition that
\begin{equation}
\lim_{x \rightarrow \infty} \frac{f(\lambda x)}{f(x)} \rightarrow \lambda^{\rho}, \; \forall \lambda > 0
\end{equation}
is regularly varying with index $\rho$ denoted by $f \in RV_{\rho}$. 
}
\end{defi}

We define the class of all regularly varying functions denoted by $\mathcal{R} = \cup_{\rho\in\mathbb{R}}RV_{\rho}$.
It is also convention to distinguish the special sub-class of functions denoted generically by $L(x)$ that are regularly varying with an index of $\rho = 0$ as follows.

\begin{cor}[Slowly Varying Tail] A function $f \in \mathcal{R}$ is slowly varying if $\rho = 0$. 
\end{cor}

So one may consider the class of severity distributions or densities that are members of this class. To provide some intuition for properties of severity models in this class of regularly varying models we note the following important features. It can be shown that for loss processes in which the severity models have a strictly non-negative support, the membership of a severity distribution in the class of regularly varying functions with tail index $\rho > 0$ implies that
\begin{equation}
\mathbb{E}\left[X^{\alpha}\right] = \begin{cases}
c, & \text{for some real constant $c<\infty$, if $\beta < \rho$};\\
\infty, & \text{if $\beta > \rho$}.
\end{cases}
\end{equation}
Analogously (setting $\beta = \rho + k$) if the right tail of a distribution $\overline{F}$ is regularly varying with an index $\rho$ then this implies that the distribution $F$ will have $(\rho + k)$-th moments which are infinite for $k > 0$. In addition the following is also known about truncated moments of severity distributions in which $\overline{F}_{X}(x)$ is regularly varying at infinity with $\rho = -\alpha$ for some $\alpha > 0$.

\begin{equation}
\begin{split}
\mathbb{E}\left[X^{\beta}\mathbb{I}_{X \leq x}\right] \sim \frac{1}{\beta - \alpha}x^{\beta}\overline{F}_X(x), \;\; \beta > \alpha.\\
\mathbb{E}\left[X^{\beta}\mathbb{I}_{X > x}\right] \sim \frac{1}{\beta - \alpha}x^{\beta}\overline{F}_X(x), \;\; \beta \leq \alpha.\\
\end{split}
\end{equation}
Clearly, these truncated moments are often of direct interest in risk management and especially in the context of capital estimation.

In addition, following properties of regularly varying functions apply specifically in the context of distributions and densities, see \cite[Theorem 1.20]{soulier2009some} and \cite{mikosch1999regular}.
 
\begin{cor}[Regularly Varying Severity Distributions]
If the severity model has a regrularly varying distribution $F \in RV_{\rho}$ with a density $f$ which is locally integrable on $[1,\infty)$ with 
\begin{equation}
F(x) = \int_{1}^{x} f(t) dt,
\end{equation}
then given the severity density $f$ is ultimately monotone, one has for $\rho \neq 0$ that $f \in RV_{\rho-1}$.
Furthermore, one can show that in the case of non-negative random variables, such as for a loss process in an LDA severity model, if the distribution is regularly varying $F \in RV_{\rho}$ with $\rho \geq 0$, then the right tail $\overline{F}(x) \in RV_{-\alpha}$.
\end{cor}

Furthermore, in OpRisk when considering severity densities and distribution functions we are working with strictly positive functions. In this regard we note that positive regularly varying functions have a unique representation detailed in Theorem \ref{Thm:KaramataRV}, see \cite{bojanic1963slowly}, \cite{bojanic1971slowly}, \cite{geluk1981n} and \cite{balkema1979extension}. This representation demonstrates an important property of such positive regularly varying functions, as it shows that the integration of regularly varying functions (tail functionals) will behave in the same manner as the integration of power functions. In addition, we note that in general the class of severity distributions and densities considered in OpRisk settings will be strictly monotonic in their tail behaviour. This is significant as it means that one will achieve uniform convergence in the limit taken in the definition of regular variation for such severity distributions.

\begin{theo}[Karamata's Representation Theorem for Regularly Varying Functions]{\label{Thm:KaramataRV} A function $f$ is a positive regularly varying function at infinity, $f \in RV_{-\rho}$, with index $-\rho$ if and only if $f$ can be represented by
\begin{equation}
f(x) = c(x)\exp\left(\int_{x_0}^x \frac{-\rho + \epsilon(t)}{t}dt \right), \; \forall x \geq x_0
\end{equation}
with $c(x) = c + o(1)$ for some $c > 0$ and $\epsilon(t) = o(1)$.
}
\end{theo}

\begin{rem} It is worth considering the intuition and relevance of the Karamata Representation Theorem. In particular it demonstrates that when integrating regularly varying functions $f \in RV_{\rho}$, one can pass the slowly varying component outside the integral as follows
\begin{equation}
F(x) = \int_{0}^{x}f(t)dt = \int_{0}^x t^{\rho}L(t)dt = L(x)\int_{0}^x t^{\rho}dt = \frac{xf(x)}{\rho + 1},
\end{equation}
see discussion in \cite[p.25]{resnick2006heavy}.
\end{rem}

\begin{rem} A consequence of this representation theorem is that every regularly varying function $f \in RV_{\rho}$ will admit a representation given by,
\begin{equation}
f(x) = x^{\rho}L(x).
\end{equation}
\end{rem}

\begin{rem} One way to understand the implications of working with severity models from this class of distributions is to consider what it tells the practitioner about integration of functions with respect to such severity distributions. Ofcourse the key component in this regard will be the influence the right tail plays in such functionals. In particular if $\rho \geq -1$ and the severity density $f \in RV_{\rho}$ then the integration of such a function satisfies 
\begin{equation}
1- \overline{F}(x) = \int_{0}^{x} f(t)dt \in RV_{\rho + 1}.
\end{equation}
Furthermore, if the index of regular variation is $\rho \leq -1$ then one has
\begin{equation}
\overline{F}(x) = \int_{0}^{x} f(t)dt \in RV_{\rho + 1}.
\end{equation}
\end{rem}

In addition a distribution $F$ which is regularly varying with index $\rho$ can be characterised by what is widely known as the \textsl{tail balance condition} in Definition \ref{Defn:TailBalance}, see \cite{jessen2006regularly}.

\begin{lem}[Tail Balance Condition]{\label{Defn:TailBalance} A distribution function $F$ is regularly varying with index $\rho \geq 0$ if there exists $p,q \geq 0$ with $p + q = 1$ and a slowly varying function $L(x)$, which for all $\lambda > 0$ means
\begin{equation}
\lim_{x \rightarrow \infty} \frac{L(x\lambda)}{L(x)} = 1,
\end{equation} 
and the following tail balance conditions are satisfied as $x \rightarrow \infty$
\begin{equation}
\begin{split}
F(-x) &= q x^{-\rho}L(x)\left(1 + o(1)\right) \\
\overline{F}(x) &= p x^{-\rho}L(x)\left(1 + o(1)\right).
\end{split}
\end{equation}
}
\end{lem}

Finally, it is important in the context of this manuscript which is considering capital approximations to make the connection between the properties of the regularly varying distribution and its inverse (quantile function) that is often utilised pointwise as the mathematical measure of OpRisk capital, such as in the Basel II/III stipulated Value-at-Risk for some quantile level $\alpha$. 

\begin{lem}[Regularly Varying Distribution and Quantile Functions]
If the right tail of a distribution is regularly varying at infinity such that $\overline{F} \in RV_{-\rho}$ with $\rho > 0$ and $\overline{F}(x) = x^{-\rho}L(x)$, then it is also true that the quantile function $F^{\leftarrow}(t) = \inf\left\{x:F(x) \geq t \right\}$ will be regularly varying at the origin, $F^{\leftarrow}(t) \in RV_{-\rho}$. If one defines $Q(t) := F^{\leftarrow}\left(1-\frac{1}{t}\right)$ on $[1,\infty)$ then this leads to the representation $ Q(t) = t^{\frac{1}{\rho}}L^{\sharp}\left(t^{\frac{1}{\rho}}\right)$, where $L^{\sharp}$ represents the De Bruyn conjugate of $L^{-\frac{1}{\rho}}$, see details in \cite[p.79]{beirlantstatistics}.
\end{lem}

Typical examples of slowly varying functions are functions converging to a positive constant, logarithms and iterated logarithms. Another important note to make here is that distributions such as the Pareto, Cauchy, Student-t, Burr and log-gamma, truncated $\alpha$-stable distributions have regularly varying tails and are ultimately infinitely differentiable and their derivatives are regularly varying.

One can also consider a related sub-classes of regularly varying severity models known as the class of Smoothly Varying functions given in Definition \ref{Defn:SmoothVarying}, see \cite{tarov2004smoothly} and \cite[p.6]{barbe2009asymptotic}.

\begin{defi}[Smoothly Varying Function]\label{Defn:SmoothVarying} A real measurable function $f \in SR_{-\rho}(m)$ is smoothly varying with index $-\rho$ and order $m$ if it is eventually (asymptotically) $m$-times continously differentiable and the $m$-th derivative $D^mf(x) = f^{(m)}(x)$ is regularly varying with index $f^{(m)} \in RV_{-\rho -m}$. Furthermore, for any non-integer value $u > 0$, with $u=m+r$ for $m \in \mathbb{N}$ and $r \in [0,1)$, then a function $f(x)$ is smoothly varying with index $-\rho$ and order $u$ if $f \in SR_{-\rho,m}$ and the following limit holds
\begin{equation}
\lim_{\delta \rightarrow 0} \limsup_{x \rightarrow \infty}\sup_{0 < |\lambda| < \delta} \frac{f^{(m)}\left(x(1-\lambda)\right) - 
f^{(m)}(x)}{|\lambda|^{r}f^{(m)}(x)} = 0.
\end{equation}
\end{defi}

One can also state the smooth variation theorem which provides asymptotic bounds on the severity density, see \cite{Bingham1989}.

\begin{theo}[Smooth Variation Theorem] If a severity density model has a regularly varying right tail $f \in RV_{\rho}$ then there exists functions $f_1$ and $f_2$ with $f_2 \in SR_{\rho}$ and asymptotic equivalence $f_1 \sim f_2$ such that in some neighbourhood of infinity the following bounds apply $f_1 \leq f \leq f_2$. 
\end{theo}

\begin{cor} If the severity density has a regularly varying right tail, $f \in RV_{\rho}$, then there exists a function $g \in SR_{\rho}$ with $g \sim f$.
\end{cor}

This further characterization of a sub-class of regularly varying severity models is highly relevant as it allows one to make some comments on integrals of tail functionals with respect to such severity distributions. In particular, one can show the following is true for two smoothly varying functions when utilised to construct a product function, see \cite[p.47]{bingham1989regular}.

\begin{lem}[Products of Smoothly Varying Functions]
Given two smoothly varying functions $f \in SR_{\rho}$ and $\phi \in SR_{\alpha}$, then the product of these two functions is also smoothly varying as $f(x)\phi(x) \in SR_{\rho + \alpha}$.
\end{lem}

\begin{rem}[Implications for Capital Approximations]
This product closure for the family of smoothly varying functions, when combined with convolution closure properties of regularly varying functions to be explained below is particularly useful when integrating tail functionals of compound processes, such as would be required when calculating capital approximations under certain risk measures such as Expected shortfall or Spectral Risk Measures.
\end{rem}

Building on these sub-classes of regularly varying functions, one can also define additional notions or tail variation such as dominantly varying, subversively varying and long tailed severity models, see discussion in \cite{bardoutsos2011characterization}

\begin{defi}[Dominantly Varying Tail]{\label{Defn:DominantlyVaying} A severity distribution function $F$ is said to have a dominantly varying tail if it satisfies the asymptotic condition that
\begin{equation}
\limsup_{x \rightarrow \infty} \frac{\overline{F}(ux)}{\overline{F}(x)} < \infty,
\end{equation}
for any $u\in(0,1)$. An alternative equivalent relationship is to consider
\begin{equation}
\liminf_{x \rightarrow \infty} \frac{\overline{F}(ux)}{\overline{F}(x)} > 0,
\end{equation}
for any $u > 1$.
}
\end{defi}

This notion of dominated variation is interesting to consider for the following reasons discussed in \cite{goldie1978subexponential}. It is well known that for a severity model with a positive support, if the distribution $F$ has a tail $\overline{F}$ which is regularly varying, then in this case it will imply that $F$ is in the family of subexponential distributions. Alternatively, if the tail of the severity distribution $\overline{F}$ is instead merely of dominated variation, then it is no longer the case that the severity distribution $F$ needs to be in the family of subexponential models.

\begin{defi}[Subversively Varying Tail]{\label{Defn:SubversivelyVaying} A severity distribution function $F$ is said to have a subversively varying tail if it satisfies the asymptotic condition that
\begin{equation}
\limsup_{x \rightarrow \infty} \frac{\overline{F}(ux)}{\overline{F}(x)} < 1,
\end{equation}
for any $u>1$. 
}
\end{defi}

\begin{defi}[Long Tailed Distribution]{\label{Defn:LongTailed} A severity distribution function $F$ is said to be long-tailed if it satisfies the asymptotic condition that
\begin{equation}
\lim_{x \rightarrow \infty} \frac{\overline{F}(x-y)}{\overline{F}(x)} = 1,
\end{equation}
for all constants $y \in \mathbb{R}$. Equivalently one can state this by saying that $\overline{F}(x-y)$ is asymptotically equivalent to $\overline{F}(x)$, that is $\overline{F}(x-y) \sim \overline{F}(x)$.
}
\end{defi}

The long tailed severity model case is particular interesting as members of this family have the property that the distribution of a random variable $X \sim F$ is said to have a long right tail if for all $\lambda > 0$,
\begin{equation}
\lim_{x \to \infty} \mathbb{P}\mathrm{r}\left[X>x+\lambda | X>x\right] =1,
\end{equation}
or equivalently put in terms of asymptotic order it means that
\begin{equation}
\overline{F}(x+\lambda) \sim \overline{F}(x).
\end{equation}

Therefore, the interpretation of a right long-tailed distributed quantity is that if the long-tailed quantity exceeds some high level, the probability approaches 1 that it will exceed any other higher level. Put simply if you know the loss amounts are significant then the realized losses from such a severity model are probably worse than you think. 

\begin{rem}
One can show that all long-tailed distributions are heavy-tailed, but the converse is false. In addition one has that all subexponential distributions are long-tailed, but examples can be constructed of long-tailed distributions that are not subexponential.
\end{rem}

Having defined these different families of tail behaviour in a severity distribution or density, we note the following relationships between these families, see discussion in \cite{geluk2009asymptotic}. 

\begin{rem}[Relating the Families of Severity Models by Tail Behaviour] 
The following relationships between the different families of severity distributions, classified by their right tail behaviour holds:
\begin{enumerate}
\item{Firstly, the class of sub-exponential distributions is larger than the class of regularly varying distributions and one can observe the relationship through the result in Lemma \ref{Lemma:SubexpRV}, see \cite[Lemma 3.2]{jessen2006regularly}. Secondly, it is well known that the class of smoothly varying functions, functions with continous derivatives being regularly varying at infinity, is a subclass of regularly varying functions.}
\item{The intersection between the family of dominantly varying, subversively varying and sub-exponential tailed distributions is contained in the family comprised of the intersection between the  dominantly varying tailed functions and the long tailed functions. Furthermore, these sub-families formed from the intersections are themselves contained in the family of sub-exponential models which is iteself contained in the family of long tailed distributions.}
\end{enumerate}
\end{rem}

It will be useful in generalizing assumptions on the frequency distribution in the LDA structure when obtaining the results for the single loss capital approximations to also consider two additional classification concepts for the families of heavy tailed severity models, these are the notions of extended regularly varying functions (ER) and the O-regularly varying functions (OR) given in Definition \ref{Def:ERandORFUnctions}. 

First we define the following properties for the severity density, using the notation \cite{Bingham1989}, the lim sup and lin inf according to Equation \ref{Eqn:Linsupf} and Equation \ref{Eqn:LinInff} for $\lambda > 0$
\begin{equation}\label{Eqn:Linsupf}
f^*(\lambda) = \limsup_{x\rightarrow \infty}\frac{f(\lambda x)}{f(x)} 
\end{equation}
and
\begin{equation}\label{Eqn:LinInff}
f_*(\lambda) = \liminf_{x\rightarrow \infty}\frac{f(\lambda x)}{f(x)}
\end{equation}
with the relationship that $f_*(\lambda) = \frac{1}{f^*(1/\lambda)}$.

\begin{defi}[Extended and O-Type Regular Variation]\label{Def:ERandORFUnctions} The class of extended regularly varying functions is the set of postive measurable functions $f \in ER$ satisfying for some constants $c,d$ the relationship
\begin{equation}
\lambda^d \leq f_*(\lambda) \leq f^*(\lambda) \leq \lambda^c, \; \forall \lambda \geq 1.
\end{equation}
The class of O-regularly varying functions is the set of postive measure functions $f \in OR$ satisfying 
\begin{equation}
0 < f_*(\lambda) \leq f^*(\lambda) < \infty, \; \forall \lambda \geq 1.
\end{equation}
\end{defi}

It will also be beneficial to recall the definition of the Matuszewska index, see \cite{matuszewska1964generalization} and \cite[page 68]{Bingham1989}.

\begin{defi}[Matuszewska Index]{\label{Def:Matuszewska} Let $f$ be a positive density function, then the upper Matuszewska index, denoted $\alpha(f)$ is given as $x \rightarrow \infty$ by the infimum of the $\alpha$ values such that there exists a constant $C = C(\alpha)$ where for each $\Lambda > 1$ one has
\begin{equation}
\frac{f(\lambda x)}{f(x)} \leq C\left\{1 + o(1)\right\} \lambda^{\alpha} \;\;  \mathrm{uniformly} \; \mathrm{in} \; \lambda \in [1,\Lambda].
\end{equation}
The lower Matuszewska index, denoted $\beta(f)$ is analogously given as $x \rightarrow \infty$ by the supremum of $\beta$ values for which for some constant $D > 0$ and for all $\Lambda > 1$ one has
\begin{equation}
\frac{f(\lambda x)}{f(x)} \geq D\left\{1 + o(1)\right\} \lambda^{\beta} \;\; \mathrm{uniformly} \; \mathrm{in} \; \lambda \in [1,\Lambda].
\end{equation}
The following relationship between Matuszewske indexs is known for positive functions $f$
\begin{equation}
\beta(f) = - \alpha\left(\frac{1}{f}\right).
\end{equation}
}
\end{defi}

One can show the following properties of the Matuszewska Index for a postive function $f$ given in Lemma \ref{Lemma:MatuszewskaIndexProp}, see \cite[page 71]{Bingham1989}.

\begin{lem}{\label{Lemma:MatuszewskaIndexProp} Consider a positive function $f$ then the following properties w.r.t. the Matuszewska Index's sign can be shown:
\begin{enumerate}
\item{$f$ has bounded increase $f \in BI$ if $\alpha(f) < \infty$.;}
\item{$f$ has bounded decrease $f \in BD$ if $\beta(f) > -\infty$.;}
\item{$f$ has positive increase $f \in PI$ if $\beta(f) > 0$.; and}
\item{$f$ has positive decrease $f \in PD$ if $\alpha(f) < 0$.}
\end{enumerate}
}
\end{lem}

Of direct relevance to the results to be discussed in this manuscript on higher order asymptotic tail expansions will be the extension of the concept of Matuszewska indices which was further developed for distribution function tails in \cite{cline1994subexponentiality} who provided the statements in Lemma \ref{Lemma:MatuszewskaIndexTails}.

\begin{lem}[Matuszewska Indices for Distribution Functions]{\label{Lemma:MatuszewskaIndexTails} Given a severity distribution $F$, then the upper Matuszewska index for the tail of the distribution $(\overline{F}(x) = 1-F(x))$ denoted by $\gamma_{\overline{F}}$ is given by
\begin{equation}
\gamma_{\overline{F}} = \inf\left\{-\frac{\log \overline{F}_*(u)}{\log u} : u > 1 \right\} = - \lim_{u \rightarrow \infty} \frac{\log \overline{F}_*(u)}{\log u}.
\end{equation}
The lower Matuszewska index is given for the tail of a distribution analogously by
\begin{equation}
\delta_{\overline{F}} = \sup\left\{-\frac{\log \overline{F}^*(u)}{\log u} : u > 1 \right\} = - \lim_{u \rightarrow \infty} \frac{\log \overline{F}^*(u)}{\log u}.
\end{equation}
Analogous definitions can also be developed for severity density functions.
}
\end{lem}

When the upper and lower Matuszewska indices are finite for the tails of a distribution function one may state the following bounds in Proposition \ref{Prop:FiniteMatuszewskaIndexTailDist}.

\begin{prop}{\label{Prop:FiniteMatuszewskaIndexTailDist} Given a severity distribution function $F$ with a finite upper Matuszewska index $\gamma_{\overline{F}}<\infty$ then there exists constants $C_1$ and $x_0$ such that the bound
\begin{equation}
\frac{\overline{F}(x)}{\overline{F}(y)} \leq C_1\left(\frac{x}{y}\right)^{-\gamma}
\end{equation}
holds for all $x \geq y \geq x_0$ and $\gamma_{\overline{F}} < \gamma < \infty$. Furthermore, if the lower Matuszewska index is finitely positive $\delta_{\overline{F}} > 0$ then there exists constants $C_2$ and $x_0$ such that
\begin{equation}
\frac{\overline{F}(x)}{\overline{F}(y)} \geq C_2\left(\frac{x}{y}\right)^{-\delta}
\end{equation}
holds for all $x \geq y \geq x_0$ and $0 < \delta < \delta_{\overline{F}}$.
}
\end{prop}

We now proceed with the process of utilizing these characterizations for the heavy-tailed severity models to obtain asymptotic bounds for the compound process tails and capital estimates.

%%%%%%%%%%%%%%%%%%%%%%%%%%%%%%%%%%%%%%%%%%%%%%%%%%%%%%%%%%%%%%%%%%%%%%%%%%%%%%%%%%%%%%%%%%%%%%%%%%
\subsection{Single Risk Closed Form Compound Process Approximations of Asymptotic Tail Behaviour}
\label{Chapter_LargeLossesPV}
%%%%%%%%%%%%%%%%%%%%%%%%%%%%%%%%%%%%%%%%%%%%%%%%%%%%%%%%%%%%%%%%%%%%%%%%%%%%%%%%%%%%%%%%%%%%%%%%%%

We begin by noting several important properties one can obtain when combining severity distributions from the above heavy-tailed families into LDA structures. We will focus purely on Poisson processes, though this can trivially be generalized to other frequency distributions of interest. We first consider the annual loss process for a fixed number of loss events $N=n$ and state some properties of the partial sum with respect to assumptions on the tail behaviour of the severity model, see discussion in \cite{wang2005closure}, \cite{asmussen2003asymptotics} and \cite{embrechts1982convolution}.

\begin{lem}[Convolution Root Closure of Sub-exponential Distributions]{\label{Lemma:SubexpRV} Assume the partial sum $Z_n = \sum_{i=1}^n X_i$ is regularly varying with index $\rho \geq 0$ with each $X_i$ being i.i.d. with positive support. Then for all $i \in \left\{1,\ldots,n\right\}$ the $X_i$'s are regularly varying also with index $\rho$ and the following asymptotic equivalence as $x \rightarrow \infty$ holds
\begin{equation}
\overline{F}_{Z_n}(x) = \mathbb{P}\mathrm{r}\left(Z_n > x\right) \sim n\mathbb{P}\mathrm{r}\left(X_1 > x\right), \; \forall n \geq 1.
\end{equation}
}
\end{lem}

This result can be restated analogously, for any $n \geq 1$, by showing that one has asymptotically as $x \rightarrow \infty$,
\begin{equation}
    \overline{F^{*n}}(x) \sim n\overline{F}(x) \quad \mbox{as } x \to \infty. 
\end{equation}
This means that given a sum of $n$ independent random variables $X_1,\ldots,X_n$ with common distribution $F$ one has the following probabilistic interpretation for sub-exponential distributed random variables (often referred to as the `big jump'),
\begin{equation}
\mathbb{P}\mathrm{r}\left[X_1+ \cdots X_n>x\right] \sim \mathbb{P}\mathrm{r}\left[\max\left(X_1, \ldots,X_n\right)>x \right], \; \mathrm{as} \; x \to \infty. 
\end{equation}

In addition, if one wishes to consider insurance mitigation in which each loss $\left\{X_i\right\}_{i=1}^n$ is mitigated by an insurance policy coverage to produce $\left\{c_i X_i\right\}_{i=1}^n$ for some $c_i \in [0,1]$, as for example under one of the insurance policy stuctures detailed in \cite{peters2011impact}. Then given the losses are from a severity model which is regularly varying and tail balanced, the result in Lemma \ref{Lemma:PaulRegVar}, see \cite[Appendix 3.26]{embrechts2011modelling} can be interpreted as a generalization of the above result, in an OpRisk context, to incorporate simple insurance mitigations.

\begin{lem}[Generalization to Insurance Mitigation]\label{Lemma:PaulRegVar} Consider an i.i.d. sequence of loss random variables $X_i$ for $i \in \left\{1,2,\ldots,n\right\}$ with distribution function $F$ which is regularly varying with index $\rho \geq 0$ and satisfies the tail balance condition for some $p+q = 1$ in Definition \ref{Defn:TailBalance}. Then for any real constants $c_i \in \mathbb{R}$ and integer $n \geq 1$ one can show
\begin{equation}
\mathbb{P}\mathrm{r}\left(c_1 X_1 + \cdots + c_n X_n > x\right) \sim \mathbb{P}\mathrm{r}\left(|X_1| > x\right) \sum_{i=1}^n \left[p\left(c_i^{+}\right)^{\rho} + q\left(c_i^+\right)^{\rho}\right]
\end{equation}
where $c_i^+ = 0 \vee +c_i$ and $c_i^- = 0 \vee -c_i$.
\end{lem}

\begin{rem}
Note that in general, except when $n = 2$, it is not true that given a linear combination $c_1 X_1 + c_2 X_2 + \cdots + c_n X_n$ which is regularly varying with index $\rho$ for an i.i.d. sequence of losses $\left(X_i\right)$ , this does not imply that $X_1$ is from a distribution which is regularly varying. 
\end{rem}

If we now consider the compound process setting, in which the number of losses in the given year is treated as a random variable, then the following results can be developed for the right tail approximation as a function of the heavy-tailed assumptions of the severity model. 

For the class of heavy tailed sub-exponential LDA models we have that a probability distribution function $F$ will belong to the sub-exponential class $\mathcal{F}$ if $F \sim_M F$, i.e. it is max-sum-equivalent with itself and that the class $\mathcal{F}$ is closed under convolutions. The implications of this for LDA models is clear when one observes that sub-exponential LDA models are compound process random sums comprised of an infinite mixture of convolved distributions,  
\begin{equation}
G(x) = \sum_{n=0}^{\infty} p_n F^{n\star}(x),
\end{equation}
for a suitable series $\left\{p_n\right\}$, (\textsl{e.g. convergent sequence satisfying Kolmogorov three series theorem}). As an example, consider the case in which $p_n$ is defined by an LDA model constructed as a compound Poisson distribution, where each term will be given by $p_n = e^{-\lambda}\frac{\lambda^n}{n!}$. Then using the property of max-sum equivalence one can show the practically relevant asymptotic equivalence between the severity distribution $F$ and the annual loss distribution $G$ in which selecting $F \in \mathcal{F}$ results in $G \in \mathcal{F}$ and 
$$
\lim_{x \rightarrow \infty} \frac{\overline{G}(x)}{\overline{F}(x)} = \lambda.
$$

This asymptotic equivalence relationship between the severity distribution $F$ and the annual loss distribution $G$, which is present for sub-exponential LDA models, greatly benefits the formulation of asymptotic approximations of tail functionals such as quantiles and tail expectations in such LDA models. It should also be noted that there also exists a special sub-family of such sub-exponential models which have the additional feature of being infinitely divisible in their severity models. The consequence of this additional feature is substantial as it often means that closed form representations of the distribution and density for the annual loss distributions can be obtained, see discussions in \cite{peters2011impact} and \cite{peters2011analytic}. 

In general based on these properties we can obtain asymptotic approximations to the annual loss distribution tails which typically fall under one of the following classifications:
\begin{itemize}
\item{``First-Order'' and ``Second-Order'' Single Loss Approximations: recently discussed in \cite{bocker2009first}, \cite{degen2010calculation}, \cite{degen2011scaling} and references therein.}
\item{``Higher-Order'' Single Loss Approximations: see discussions in \cite{bingham1989regular} and recent summaries in \cite{albrecher2010higher} and references therein.}
\item{Extreme Value Theory (EVT) Single Loss Approximations (Penultimate Approximations): the EVT based asymptotic estimators for \underline{linear normalized} and \underline{power normalized} extreme value domains of attraction were recently discussed in \cite{degen2011scaling}.}
\item{Doubly Infinitely Divisible Tail Asymptotics given $\alpha$-stable severity models discussed in \cite{peters2011analytic} and \cite{peters2011impact}}
\end{itemize}

%%%%%%%%%%%%%%%%%%%%%%%%%%%%%%%%%%%%%%%%%%%%%%%%%%%%%%%%%%%%%%%%%%%%%%%%%%%%%%%%%%%%%%%%%%%%%%%
\subsection{First Order Single Risk Loss Process Asymptotics for Heavy-Tailed LDA Models}
%%%%%%%%%%%%%%%%%%%%%%%%%%%%%%%%%%%%%%%%%%%%%%%%%%%%%%%%%%%%%%%%%%%%%%%%%%%%%%%%%%%%%%%%%%%%%%%

Consider the compound process distribution and right tail distribution functions, given for annual loss $Z_N = \sum_{i=1}^N X_i$, by
\begin{equation*}
\begin{split}
F_{Z_N}(x) &= \sum_{n=0}^{\infty} p_n F^{*n}(x) \\
\overline{F_{Z_N}}(x) &= \sum_{n=1}^{\infty} p_n \overline{F^{*n}}(x),
\end{split}
\end{equation*}
where it is assumed that the frequency probability mass function satisfies $\sum_{n=0}^{\infty}np_n < \infty$. Then the following first order single risk loss processes tail asymptotic results apply when one considers a severity model with a regularly varying right tail.

\begin{theo}[First Order Single Risk Asymptotic Tail Approximation]\label{Prop:RandSumResults} The following results for compound process tail behaviour hold:
\begin{enumerate}
\item{Assuming the distribution of $X_i$ losses is in the subexponential severity family and an independent integer-valued random variable for the number of losses, $N$ satisfies $\mathbb{E}\left[(1+\epsilon)^N \right] < \infty,$ for an $\epsilon > 0$. Then the following asymptotic equivalence holds
\begin{equation}
\mathbb{P}\mathrm{r}\left(Z_N > x\right) \sim \mathbb{E}[N]\mathbb{P}\mathrm{r}\left(X_1 > x\right).
\end{equation}
}
\item{Assuming the distribution of $X_i$ is regulary varying with postive index $\rho>0$ and the mean number of losses is finite $\mathbb{E}[N] < \infty$ and $\mathbb{P}\mathrm{r}\left(N>x \right) = o\left(\mathbb{P}\mathrm{r}\left(X_1 > x\right)\right)$. Then the following asymptotic equivalence holds
\begin{equation}
\mathbb{P}\mathrm{r}\left(Z_N > x\right) \sim \mathbb{E}[N]\mathbb{P}\mathrm{r}\left(X_1 > x\right).
\end{equation}
}
\item{If one assumes the counting random varible $N$ is regularly varying with index $\beta \geq 0$. If $\beta = 1$ then assume that $\mathbb{E}[N] < \infty$. Let the i.i.d. loss random variables $(X_i)$ with finite mean $\mathbb{E}\left[X_i\right] < \infty$ and tail probablity with asymptotic condition $\mathbb{P}\mathrm{r}\left( X_i > x\right) = o\left(\mathbb{P}\mathrm{r}\left(N > x\right)\right)$, then assymptotically one has the equivalence for the compound processes
\begin{equation}
\mathbb{P}\mathrm{r}\left(Z_N > x\right) \sim \left(\mathbb{E}[X_1]\right)^{\beta}\mathbb{P}\mathrm{r}\left(N > x\right).
\end{equation}
}
\item{Assume that $\mathbb{P}\mathrm{r}\left(N>x\right) \sim c \mathbb{P}\mathrm{r}\left(X_1 > x\right)$ for some $c > 0$ and $X_1$ is regularly varying with an index $\rho \geq 1$ and with finite mean $\mathbb{E}\left[X_1\right] < \infty$. Then one has the asymptotic equivalence
\begin{equation}
\mathbb{P}\mathrm{r}\left(Z_N > x\right) \sim \mathbb{P}\mathrm{r}\left(X_1 > x\right) \left(\mathbb{E}[N] + c\left(\mathbb{E}\left[X_1 \right] \right)^{\rho} \right).
\end{equation}
}
\end{enumerate}

\end{theo}

The above results make intuitive sense, since in for example the heavy-tailed subexponential severity model case, the tail of the distribution of the sum and the tail of the maximum are asymptotically of the same order. Hence the sum of losses is governed by large individual losses, and therefore the distributional tail of the sum is mainly determined by the tail of the maximum loss. In addition in the case of the general sub-exponential models, we note that the results presented for the first order single loss tail approximation required that $\mathbb{E}[z^N] < \infty$ for some $z > 1$. This means that $N$ must have finite moments of all orders. This clearly has implications on the properties of the severity distribution since we observed the result that 
\begin{equation} \label{eqn:FirstOrderSLA}
\lim_{x \rightarrow \infty} \frac{\overline{F_{Z_N}}(x)}{\overline{F}(x)} =\mathbb{E}[N].
\end{equation}
To understand these implications we shall present a basic understanding how this first order asymptotic is derived.

For these first order asymptotic behaviours of the right tail $\overline{F_{Z_N}}(x)$ one can consider obtaining an upper bound for the asymptotic ratio of the tail of the compound distribution and the severity distribution tail for each number of losses $n \in \mathbb{J}$. This is given in Lemma \ref{Lemma:KestenBound} for the standard geometric Kesten bound and then in the more general bound in Theorem \ref{Thm:BoundGeneral} derived in \cite[Theorem 7]{daley2007tail}.

\begin{lem}[Kesten's Bound]\label{Lemma:KestenBound} If the severity distribution $F$ is in the class of sub-exponential distributions then for each $\epsilon > 0$ there exists a constant $K = K(\epsilon) < \infty$ such that for all $n \geq 2$ the following bound holds
\begin{equation}
\frac{\overline{F^{*n}}(x)}{\overline{F}(x)} \leq K(1+\epsilon)^n, \; x \geq 0.
\end{equation}
\end{lem}

A generalized version of such an upper bound for the tail ratios for each number of losses $n$ of $\frac{\overline{F^{*n}(x)}}{\overline{F}(x)}$ is given in Theorem \ref{Thm:BoundGeneral}.

\begin{theo}{\label{Thm:BoundGeneral} Consider a severity distribution $F$ which is subexponential. Next define the sequence of constants $\left\{\alpha_n\right\}_{n \geq 0}$ given by
\begin{equation}
\alpha_n = \sup_{x \geq 0} \frac{\overline{F^{*n}}(x)}{\overline{F}(x)}.
\end{equation}
Then a general bound on $\alpha_n$ for each $n$ which is only a function of $\overline{F}$ and $\overline{F^{*2}}$ is obtained by using for any $n = 1,2,3,\ldots$
\begin{equation}
\alpha_n \leq \sum_{j=1}^{n-1}\left(1+c_j\right)\prod_{k=j+1}^{n-1}(1+\epsilon_k) + \prod_{k=1}^{n-1}(1+\epsilon_k)
\end{equation}
with the empty product $\prod_{k=1}^{n-1}(1+\epsilon_k) = 1$ and where the sequence $T_n$, $\epsilon_n$ are defined to satisfy
\begin{equation}
\sup_{x \geq T_n} \frac{\overline{F^{*2}}(x)}{\overline{F}(x)} - 1 \leq 1 + \epsilon_n 
\end{equation}
where $\epsilon_n$ can be selected arbitrarily small through selection of $T_n$ large and $c_n$ is given by
\begin{equation}
c_n = \frac{F(T_n)}{\overline{F}(T_n)}.
\end{equation}
}
\end{theo}

Then using one of these bounds on the tail ratios of the n-fold convolution of the severity model and the tail of the marginal severity model for each $n$ one can extend to the compound process setting using the dominated convergence theorem.
Given for example the standard Kesten bound one now applies the dominated convergence theorem given in Lemma \ref{Lemma:DomConv} to the limit to obtain the result for the First Order Single Loss Approximation in (\ref{eqn:FirstOrderSLA}). 

\begin{lem}[Dominated Convergence Theorem] \label{Lemma:DomConv} Consider a sequence of integrable functions $\left\{f_n\right\}$ on probability space $(\Omega,\Sigma,\mu)$ which satisfy the limit that
\begin{equation}
\lim_{n \rightarrow \infty} f_n(x) = f(x), \; \mathrm{almost} \; \mathrm{everywhere} - \mu.
\end{equation}
In addition suppose that there exists an integrable function $$H \geq 0: \left|f_n(x)\right| \leq H(x) \; \forall n$$ 
then if $|f(x)| \leq H(x)$ one has the limit result
\begin{equation}
\lim_{n \rightarrow \infty} \int_{\Omega} f_n(x) d\mu(x) = \int_{\Omega} f(x) d\mu(x).
\end{equation}
\end{lem}

Therefore one utilise this bound and the dominated convergence theorem to interchange the order of the summation and the limit and then utilise the fact that for heavy-tailed sub-exponential severity models the condition that
\begin{equation}
\lim_{x \rightarrow \infty}\frac{\overline{F^{*2}}(x)}{\overline{F}(x)} = 2
\end{equation}
characterizing this sub-class of distributions implies that 
\begin{equation}
\lim_{x \rightarrow \infty}\frac{\overline{F^{*n}}(x)}{\overline{F}(x)} = n
\end{equation}
see \cite[Lemma 1.3.4]{embrechts1997modelling}. Hence one obtains for ths compound process ratio,
\begin{equation}
\lim_{x \rightarrow \infty} \frac{\overline{F_{Z_N}}(x)}{\overline{F}(x)} = \lim_{x \rightarrow \infty} \sum_{n=1}^{\infty} p_n \frac{\overline{F^{*n}}(x)}{\overline{F}(x)} = \sum_{n=1}^{\infty} n p_n =\mathbb{E}[N],
\end{equation}
which is equivalent to the asymptotic equivalence statement that
\begin{equation}
\overline{F_{Z_N}}(x) \sim \mathbb{E}[N]\overline{F}(x).
\end{equation}

\begin{rem}
In \cite{daley2007tail} they weaken the condition of all moments of $N$ being finite, based on the results presented in \cite{stam1973regular}. This relaxation involves considering less restrictive tail assumptions on the sub-exponential severity distribution $\overline{F}$, in particular it is assumed instead that it is a function of O-regular variation $\overline{F} \in OR$. Then two bounds are obtained for the ratio $\frac{\overline{F^{*n}}(x)}{\overline{F}(x)}$ depending on whether the lower Matuszewska index $\beta(\overline{F})$ is in the interval $(-\infty,-1]$ or $(-1,0)$ then the corresponding bounds can be obtained where
\begin{equation}
\frac{\overline{F^{*n}}(x)}{\overline{F}(x)} \leq A n^{|\beta(\overline{F})|+1+\varepsilon}, \; \mathrm{for} \; \overline{F} \in OR \; \mathrm{and} \; \beta(\overline{F}) < -1
\end{equation}
for some constants $A$ and $\varepsilon > 0$ or 
\begin{equation}
\frac{\overline{F^{*n}}(x)}{\overline{F}(x)} \leq A n, \; \mathrm{for} \; \overline{F} \in OR \; \mathrm{and} \; \beta(\overline{F}) \in (-1,0)
\end{equation}
for some constant $A$.
\end{rem}

A summary of results is provided in \cite[Theorem 2]{daley2007tail}. 

%%%%%%%%%%%%%%%%%%%%%%%%%%%%%%%%%%%%%%%%%%%%%%%%%%%%%%%%%%%%%%%%%%%%%%%%%%%%%%%%%%%%%%%%%%%%%%%
\subsection{Refinements and Second Order Single Risk Loss Process Asymptotics for Heavy-Tailed LDA Models}
%%%%%%%%%%%%%%%%%%%%%%%%%%%%%%%%%%%%%%%%%%%%%%%%%%%%%%%%%%%%%%%%%%%%%%%%%%%%%%%%%%%%%%%%%%%%%%%
The second order single loss approximation as discussed in \cite{Albrecher2010} and derived in \cite{omey1986second} takes the form given by the Theorem \ref{Propsn:AnnualLossRefineApprox}, also see \cite{Degen2010} [Proposition A3].

\begin{theo}[Second Order Single Risk Asymptotic Tail Approximation]\label{Propsn:AnnualLossRefineApprox} The following results for compound process tial behaviour hold under a refined second order approximation:
\begin{enumerate}
\item{Assuming the severity distribution for losses $X_i$ is zero at the origin ($x = 0$) and satisfies that both the tail distribution $\overline{F}$ and the density $f$ are subexponential. Furthermore assume the the frequency distribution $N \sim F_N(n)$ is such that its probability generating function given by
\begin{equation}
p_N(v) = \mathbb{E}\left[v^N\right] = \sum_{n=0}^{\infty} \mathbb{P}\mathrm{r}(N=n)v^n, 
\end{equation}
is analytic at $v=1$, then one has for \underline{finite mean severity models} $(\mathbb{E}[X] < \infty)$, that
\begin{equation}
\lim_{x \rightarrow \infty} \frac{\overline{F}_Z(x) - \mathbb{E}[N]\overline{F}(x)}{f(x)} = \mathbb{E}[X]\mathbb{E}[(N-1)N].
\end{equation}
If the severity distribution has an \underline{infinite mean} and the density satisfies $f \in RV_{-1/\beta-1}$ for $1 \leq \beta < \infty$ then one has
\begin{equation}
\lim_{x \rightarrow \infty} \frac{\overline{F}_Z(x) - \mathbb{E}[N]\overline{F}(x)}{f(x)\int_{0}^x \overline{F}(s)ds} = c_{\beta}\mathbb{E}[(N-1)N].
\end{equation}
with $c_{1} = 1$ and $c_{\beta} = (1-\beta) \frac{\Gamma^2(1-1/\beta)}{2\Gamma(1-2/\beta)}$ for $\beta \in (1,\infty)$.
}
\item{Assuming the losses $X_i$ have a subexponential severity distribution function $F$ in which $\mathbb{E}[X] < \infty$. Furthermore, assuming that the severity distribution admits admits a continous long-tailed severity density $f(x)$ which is dominantly varying with an upper Matuszewska index given by $\alpha(f) < -1$. Futhermore, assuming that there is also an independent integer-valued random variable for the number of losses, $N$ satisfying $\mathbb{E}\left[(1+\epsilon)^N \right] < \infty,$ for an $\epsilon > 0$. Then the second-order asymptotic approximation of the compound process tail probabilities, as $x \rightarrow \infty$, are given by
\begin{equation}
\begin{split}
\overline{F}_{Z}(x) = \mathbb{E}\left[N\right]\overline{F}(x) + 2 \mathbb{E}\left[\frac{N!}{(N-2)! 2!}\right] \mathbb{E}\left[X\right]f(x).
\end{split}
\end{equation}
}
\item{Assuming the losses $X_i$ have a severity distribution function $F$ which has regularly varying tail with index $\rho \in (0,1]$ such that $\overline{F}(x) \in RV_{-\rho}$ and a severity density $f(x)$ which is regularly varying, then the second order approximation for $\overline{F}_Z(x)$ is given by
\begin{equation}
\begin{split}
\overline{F}_{Z}(x) = \begin{cases}
\mathbb{E}\left[N\right]\overline{F}(x) - \frac{(2-\rho)\Gamma(2-\rho)}{(\rho - 1)\Gamma(3-2\rho)}\mathbb{E}\left[ \frac{N!}{(N-2)!2!} \right]f(x) \int_{0}^{x}\overline{F}(y)dy, & 0 < \rho < 1 \\
\mathbb{E}\left[N\right]\overline{F}(x) + 2\mathbb{E}\left[ \frac{N!}{(N-2)!2!} \right]f(x) \int_{0}^{x}\overline{F}(y)dy, &  \rho = 1. \\
\end{cases}
\end{split}
\end{equation}
}
\end{enumerate}
\end{theo}

\begin{rem} Note many distributions will satisfy the conditions in Theorem \ref{Propsn:AnnualLossRefineApprox} such as Log-Normal, Weibull, Benktander Type I and Type II and numerous others.
\end{rem}

\begin{rem}{These first and second order repesentations are remarkable as they each state that for the annual loss distribution under an OpRisk LDA model for general sub-exponential family severity models, when considered at high confidence levels, the resulting quantiles of the annual loss distribution become only dependent on the tail of the severity distribution and not on the body. Therefore when making such an asympototic approximation it is convenient that one only requies a quantification of the mean of the frequency distribution. Consequently, over-dispersion as captured by Negative-Binomial processes will not affect the high confidence level quantiles of the annual loss distribution.}
\end{rem}

%%%%%%%%%%%%%%%%%%%%%%%%%%%%%%%%%%%%%%%%%%%%%%%%%%%%%%%%%%%%%%%%%%%%%%%%%%%%%%%%%%%%%%%%%%%%%%%
%%%%%%%%%%%%%%%%%%%%%%%%%%%%%%%%%%%%%%%%%%%%%%%%%%%%%%%%%%%%%%%%%%%%%%%%%%%%%%%%%%%%%%%%%%%%%%%
\section{The Journey Completes: Going from Compound Process Tail Asymptotics to Capital Approximations}
%%%%%%%%%%%%%%%%%%%%%%%%%%%%%%%%%%%%%%%%%%%%%%%%%%%%%%%%%%%%%%%%%%%%%%%%%%%%%%%%%%%%%%%%%%%%%%%
%%%%%%%%%%%%%%%%%%%%%%%%%%%%%%%%%%%%%%%%%%%%%%%%%%%%%%%%%%%%%%%%%%%%%%%%%%%%%%%%%%%%%%%%%%%%%%%
In this section we will consider asymptotic approximations of key risk management quantities known as risk measures which are used in the allocation of capital and reserving in all financial institutions and stipulated as standards under regulatory accords in both Basel II/III and Solvency II. Examples of such tail functionals include the calculation of Value-at-Risk (VaR), Expected Shortfall (ES) and Spectral Risk Measures as detailed below in both their definitions and the resulting simple first and second order asymptotic approximations one may consider. 

\begin{prop}[First-Order Approximate Risk Measures] These asymptotic expansions allow one to obtain estimates of common risk measures, see \cite{artzner1999coherent} and \cite{mcneil2005quantitative}, such as:
\begin{enumerate}
\item{\textbf{Value-at-Risk (VaR):} for a level $\alpha \in (0,1)$, given by the quantile of the annual loss distribution,
\begin{equation}
\begin{split}
\mathrm{VaR}_{Z}\left(\alpha\right) &= F^{\leftarrow}_{Z}(\alpha) = \inf\left\{z \in \mathbb{R}:F_{Z}(z) \geq \alpha \right\}\\
&\approx F_Z^{\leftarrow}\left(1-\frac{1-\alpha}{\mathbb{E}[N]}\left[1 + o(1) \right]\right) \sim F^{\leftarrow}\left(1-\frac{1-\alpha}{\mathbb{E}[N]}\right),
\end{split}
\end{equation}
where $F^{\leftarrow}(\cdot)$ is the generalized inverse, see \cite{embrechts2010note}.}
\item{\textbf{Expected Shortfall (ES):} for a level $\alpha \in (0,1)$, the expected shortfall (ES) is given by the tail expectation of the annual loss distribution according to
\begin{equation}
\begin{split}
\mathrm{ES}_{Z}(\alpha) &= \mathbb{E}\left[Z|Z \geq \mathrm{VaR}_{Z}\left(\alpha\right)\right] = \frac{1}{1-\alpha}\int_{\alpha}^1 \mathrm{VaR}_Z(s)ds\\
&\approx \frac{\alpha}{\alpha - 1}F^{\leftarrow}\left(1-\frac{1-\alpha}{\mathbb{E}[N]}\right) \sim \frac{\alpha}{\alpha - 1}VaR_{Z}\left(\alpha\right),
\end{split}
\end{equation}
see \cite{biagini2009asymptotics}; and}
\item{\textbf{Spectral Risk Measure (SRM):} for a weight function $\phi:[0,1] \mapsto \mathbb{R}$ given by 
\begin{equation}
\begin{split}
\mathrm{SRM}_{Z}(\phi) &= \int_{0}^1 \phi(s) \mathrm{VaR}_Z(s)ds\\
&\approx \mathcal{K}(\alpha,\phi_1)F^{\leftarrow}\left(1-\frac{1-\alpha}{\mathbb{E}[N]}\right) \sim  \mathcal{K}(\alpha,\phi_1)VaR_{Z}\left(\alpha\right),
\end{split}
\end{equation}
with $\forall t \in (1,\infty)$ a function $\phi_1(1-1/t) \leq K t^{-1/\beta + 1 - \epsilon}$ for some $K>0$ and $\epsilon > 0$ where
\begin{equation} \label{eqn:SRM_K}
\mathcal{K}(\alpha,\phi_1) = \int_{1}^{\infty} s^{1/\beta - 2}\phi_1(1-1/s)ds.
\end{equation}
}
\end{enumerate}
\end{prop}

For the Spectral Risk Measure (SRM), in \cite{tong2012asymptotics} it is shown that if an individual has a Constant Absolute Risk Aversion (CARA) utility function with coefficient of absolute risk aversion $\xi$ then the SPR should be given as
$$SRM_{\phi_\kappa}(\kappa) = \int_{\kappa}^1\phi_\kappa(s)VaR(s),$$
where 
$$\phi_\kappa(s) = (1-\kappa)^{-1}\phi_1\left( 1- \frac{1-s}{1-\kappa}\right)I_{[\kappa, 1]}(s)$$
with
$$\phi_1(\kappa)= \frac{\xi e^{-\xi(1-\kappa)}}{1-e^{-\xi}}.$$
Note that if one take $\phi_1(t)\equiv 1 \forall t \in [0,1]$ then the Spectral Risk Measure resumes to the Expected Shortfall. 
%====================================
\begin{prop}[Second-Order Approximation of the VaR] 
	Assume that severity distribution $F$ has finite mean, and the hazard rate $h(x) = \frac{f(x)}{1-F(x)}$ is of regular variation $h \in RV_{-\beta}$ for $\beta \geq 0$, then as $\alpha \rightarrow 1$ one has for the inverse of the annual loss distribution the result (see \cite{albrecher2010higher} and \cite{Degen2010}),
	\begin{equation}
		VaR_Z(\alpha) = F^{-1}_{Z}(\alpha) = F^{-1}\left(1 - \frac{1 - \alpha}{\mathbb{E}[N]}\left\{1 + \widetilde{c}_{\beta}g_1\left(F^{-1}(\widetilde{\alpha})\right) + o\left(g_1\left(F^{-1}(\widetilde{\alpha})\right) \right) \right\}^{-1} \right)
	\end{equation}
	where $\widetilde{\alpha} = 1 - (1-\alpha)/\mathbb{E}[N]$ and
	\begin{equation*}
		\begin{split}
			g_1(x) &= \begin{cases}
			\frac{f(x)}{1-F(x)}, & \mathrm{if} \, \mathbb{E}[X] < \infty,\\
			\frac{\int_{0}^x\overline{F}(s)ds f(x)}{1 - F(x)}, & \mathrm{if} \, \mathbb{E}[X] = \infty.;
			\end{cases}\\
			\widetilde{c}_{\beta} &= \begin{cases}
			\frac{\mathbb{E}[X]\mathbb{E}[(N-1)N]}{\mathbb{E}[N]}, & \mathrm{if} \, \mathbb{E}[N] < \infty,\\
			\frac{c_{\beta}\mathbb{E}[(N-1)N]}{\mathbb{E}[N]}, & \mathrm{if} \, \mathbb{E}[N] = \infty,\\
								\end{cases}
		\end{split}
	\end{equation*}
	where we define
	\begin{equation}
	c_{\beta} = \begin{cases}
	1, & \beta = 1,\\
	(1-\beta)\frac{\Gamma^2\left(1 - \frac{1}{\beta}\right)}{2\Gamma\left(1 - \frac{2}{\beta}\right)}, & \text{ if } \beta \in (1,\infty).
	\end{cases}
	\end{equation}
\end{prop}

\begin{prop}[Second-Order Approximation of the SRM] 
For a given tail function $\bar{F}(t) \in RV_{-\alpha}$ define $U_F(t):=(1/(1-F))^{\leftarrow}(t) \in RV_{1/\alpha}$. Now assume that for some $\rho \leq 0, \ \exists A(t) \in RV_\rho$ such that
	$$\lim_{t\rightarrow +\infty} \frac{1}{A(t)}\left( \frac{U_F(tx)}{U_F(t)} - x^{1/\alpha} \right) = x^{1/\alpha}\frac{x^\rho -1}{\rho}.$$
	Then, 
	$$SRM_{\phi_\kappa}(\kappa) \sim \left[ \mathcal{K}(\alpha,\phi_1) + A\left( \frac{\mathbb{E}[N]}{1-\kappa}\right) \mathcal{M}(\alpha, \phi_1, \rho) \right]F^{\leftarrow}\left(1-\frac{1-\alpha}{\mathbb{E}[N]}\right),$$
as $\kappa \rightarrow 1$, where
$$\mathcal{M}(\alpha, \phi_1, \rho) = \frac{1}{\rho}\int_1^{+\infty} t^{1/\alpha-2}(t^\rho-1)\phi_1\left( 1-\frac{1}{t}\right)dt$$
and $\mathcal{K}(\alpha,\phi_1)$ as defined in \ref{eqn:SRM_K}, see details in \cite{tong2012asymptotics}.
\end{prop}

\begin{cor}[Second-Order Approximation of the ES] 
Let $\phi_1(t)\equiv 1 \forall t \in [0,1]$. Then, when $\kappa \rightarrow 1$ we have the following Second-Order approximation of $ES(\kappa)$, see \cite{tong2012asymptotics}:
$$ES(\kappa) \sim \left[ \frac{\alpha}{\alpha -1} + \frac{\alpha^2}{(\alpha - \alpha\rho -1)(\alpha-1)} A\left( \frac{\mathbb{E}[N]}{1-\kappa}\right)\right]F^{\leftarrow}\left(1-\frac{1-\alpha}{\mathbb{E}[N]}\right).$$
\end{cor}

The properties of such asymptotic single loss approximation estimates are still an active subject of study with regard to aspects such as explicit approximation errors, unbiased quantile function estimation, asymptotic rates of convergence, sensitivity to parameter estimation and model misspecification. In the following examples, we illustrate the properties of these first and second order approximations presented above for two popular heavy tailed LDA risk models.

%%%%%%%%%%%%%%%%%%%%%%%%%%%%%%%%%%%%%%%%%%%%%%%%%%%%%%%%%%%%%%%%%%%%%%%%%%%%%%%%%%%%%%%%%%%%%%%
%%%%%%%%%%%%%%%%%%%%%%%%%%%%%%%%%%%%%%%%%%%%%%%%%%%%%%%%%%%%%%%%%%%%%%%%%%%%%%%%%%%%%%%%%%%%%%%
\section{Examples}
%%%%%%%%%%%%%%%%%%%%%%%%%%%%%%%%%%%%%%%%%%%%%%%%%%%%%%%%%%%%%%%%%%%%%%%%%%%%%%%%%%%%%%%%%%%%%%%
%%%%%%%%%%%%%%%%%%%%%%%%%%%%%%%%%%%%%%%%%%%%%%%%%%%%%%%%%%%%%%%%%%%%%%%%%%%%%%%%%%%%%%%%%%%%%%%
In this section we consider two popular examples of LDA models that are regularly considered in OpRisk settings. These are the Poisson-Log Normal and the Poisson-Inverse-Gaussian LDA models. We detail the asymptotic approximations as a function of the quantile level $\alpha$ for the VaR in both models and the tail function in the Poisson-Inverse-Gaussian, where the tail is known in closed form. In particular we consider a range of different parameter settings affecting the heavy-tailed features of the severity model and assess the accuracy of the asymptotic approximations, versus an exhaustive Monte Carlo simulation approximation of the true solutions. This provides interesting information of relevance to assessing the rates of convergence of these asymptotic results, which are currently unknown in the literature. In addition, we study the sensitivity to the parameter estimation in the accuracy of the asymptotic risk measure approximations.

%%%%%%%%%%%%%%%%%%%%%%%%%%%%%%%%%%%%%%%%%%%%%%%%%%%%%%%%%%%%%%%%%%%%%%%%%%%%%%%%%%%%%%%%%%%%%%%
\begin{exam}[Single Risk LDA Poisson-Log-Normal Family]{
Consider the heavy tailed severity model, selected to model the sequence of i.i.d. losses in each year $t$, denoted $\left\{X_i(t)\right\}_{i=1:N_t}$, and chosen to be a Log-Normal distribution $X_i \sim LN(\mu, \sigma)$ where the two parameters in this model correspond to parameterizing the shape of the distribution for the severity $\sigma$ and the log-scale of the distribution $\mu$. The survival and quantile functions of the severity are given by
\begin{equation*}
\begin{split}
&f(x;\mu,\sigma) = \frac{1}{x\sqrt{2\pi\sigma^2}}\, e^{-\frac{\left(\ln x-\mu\right)^2}{2\sigma^2}} , \; x>0; \; \mu \in \mathbb{R} \; \sigma >0 \\
&\overline{F}(x;\mu,\sigma) = 1-F(x)= \int_{x}^{\infty}\frac{1}{\sqrt{2 \pi \sigma u}}\exp\left( -\frac{1}{2 \sigma^2}\left( \log(u) - \mu^2\right) \right) du\\
& \hspace{1.6cm} =\frac12 + \frac12\,\mathrm{erf}\Big[\frac{\ln x-\mu}{\sqrt{2\sigma^2}}\Big] , \; x>0; \; \mu \in \mathbb{R} \; \sigma >0 \\
&Q(p) = \exp\left(\mu + \sigma \Phi^{-1}(p) \right), \; 0<p<1.
\end{split}
\end{equation*}
Therefore the closed form SLA for the VaR risk measure at level $\alpha$ would be presented in this case under first and second order approximations for the annual loss $Z = \sum_{n=1}^N X_i$ according to Equations (\ref{Eqn:VaRLNSLA1}) and (\ref{Eqn:VaRLNSLA2}), respectively
\begin{align} 
	\mathrm{VaR}_{\alpha}\left[Z\right] &= \exp\left[\mu - \sigma \Phi^{-1}\left(\frac{1-\alpha}{\lambda}\right) \right] \label{Eqn:VaRLNSLA1} \\
	\mathrm{VaR}_{\alpha}\left[Z\right] &= F^{-1}\left( 1- \frac{1- \alpha}{\lambda}\{ 1+\widetilde{c}_\beta g_1(F^{-1}(\widetilde{\alpha}))  \}^{-1}\right) \label{Eqn:VaRLNSLA2}
\end{align}
where $\widetilde{\alpha} = 1- (1-\alpha)/\lambda $, $g_1(x) = \frac{f(x)}{1-F(x)}$ and $\widetilde{c}_\beta = e^{\mu+\sigma^2/2}\lambda$.

We will now compare this first order and second order asymptotic result to the crude Monte Carlo approach (for which one can generate uncertainty measures such as confidence intervals in the point estimator).  To complete this example, we illustrate the basic Monte Carlo solution for the VaR for a range of quantile levels of the annual loss distribution, we display these along with the measured confidence intervals in the point estimators and we compare these to the first order SLA asymptotic result. The quantiles $\alpha \in \left\{0.70, 0.75, 0.80, 0.85, 0.9, 0.95, 0.99, 0.995, 0.9995\right\}$ are considered where the $99.5\%$ and $99.95\%$ quantile levels do in fact correspond to regulatory standards of reporting in Basel II/III.

\begin{figure}[h]%
	\centering
	\begin{minipage}{8cm}%
		\includegraphics[width=\textwidth, height=8cm]{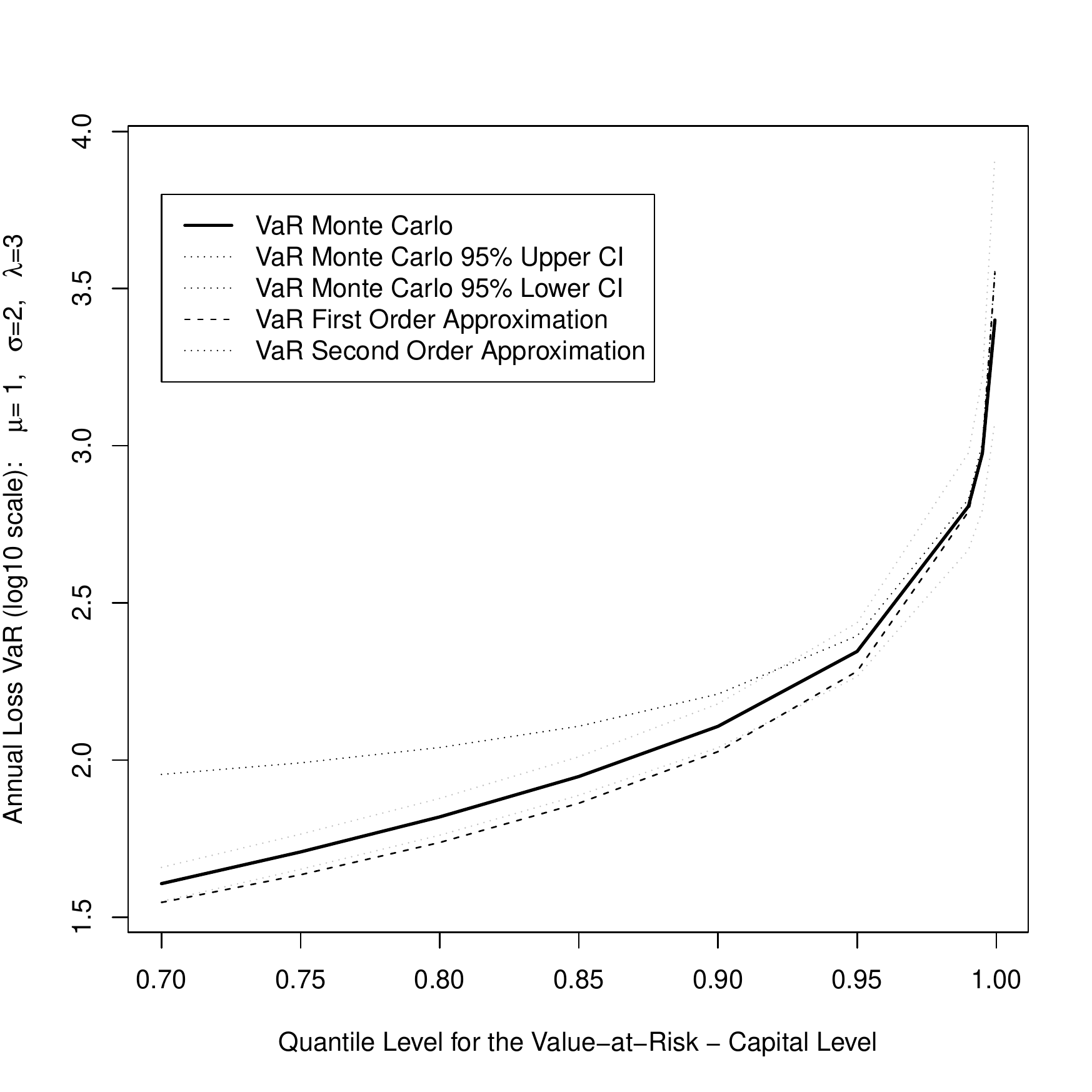}
	\end{minipage}%
	\qquad
	\begin{minipage}{8cm}%
		\includegraphics[width=\textwidth, height=8cm]{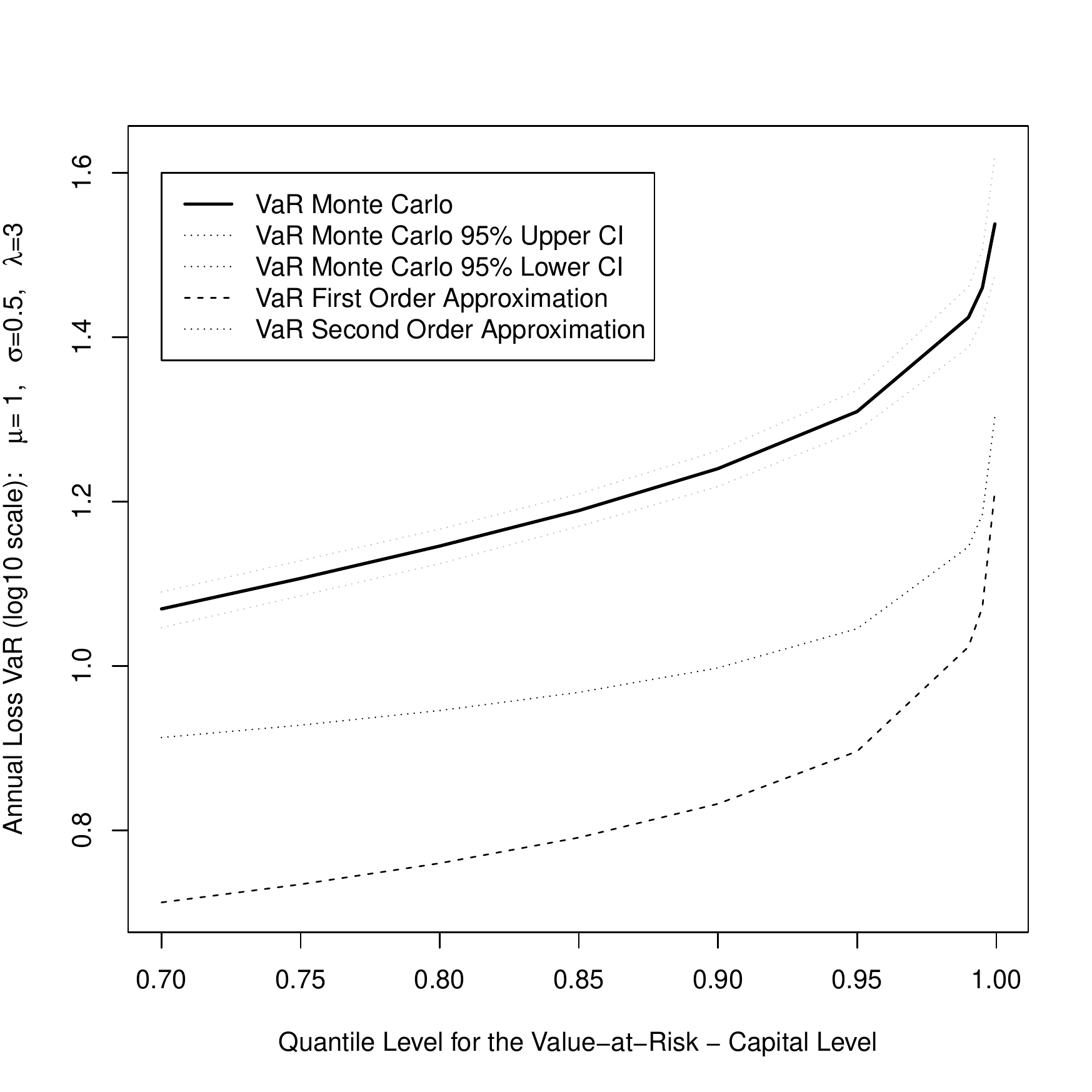}
	\end{minipage}%
	\qquad
	\begin{minipage}{8cm}%
		\includegraphics[width=\textwidth, height=8cm]{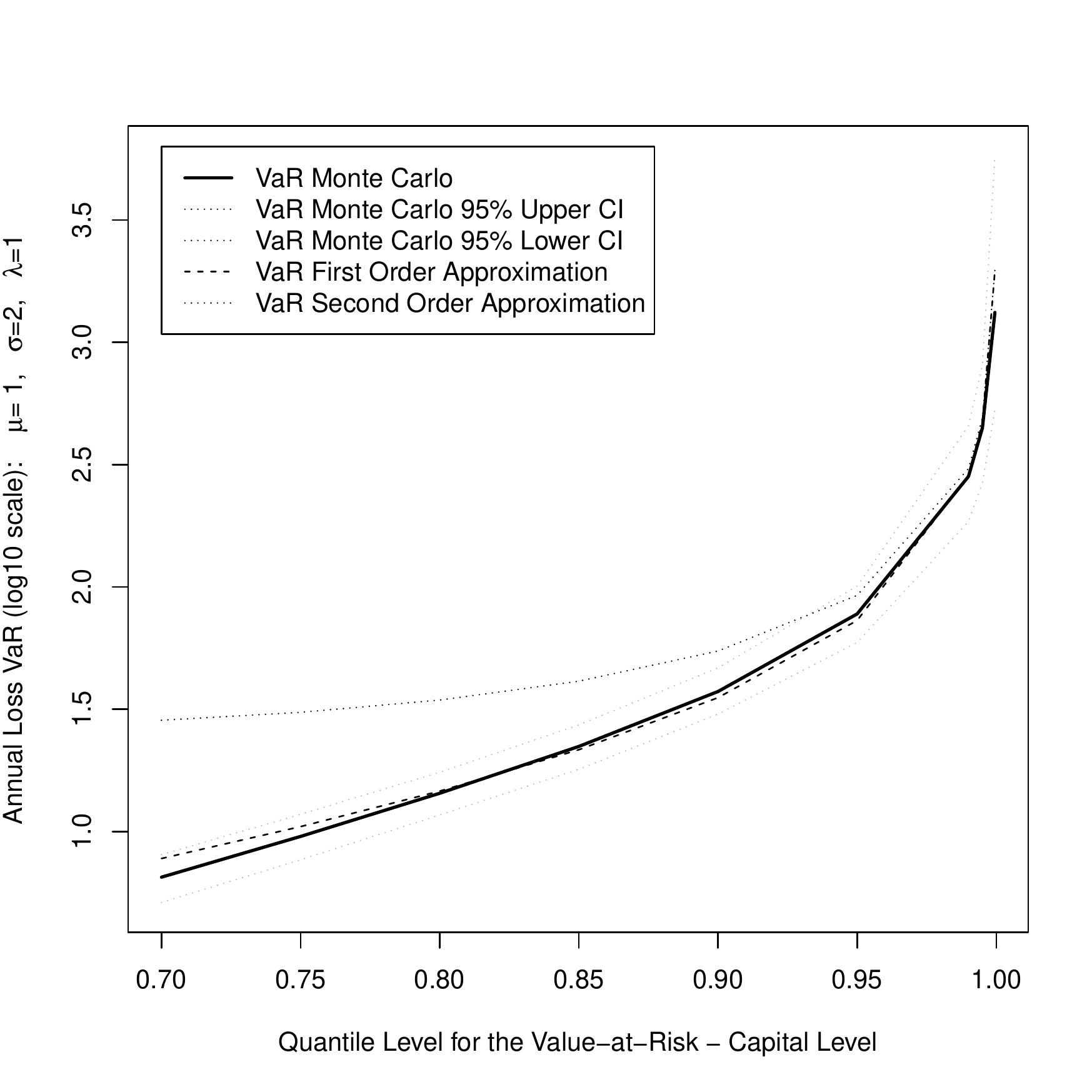}
	\end{minipage}%
	\qquad
	\begin{minipage}{8cm}%
		\includegraphics[width=\textwidth, height=8cm]{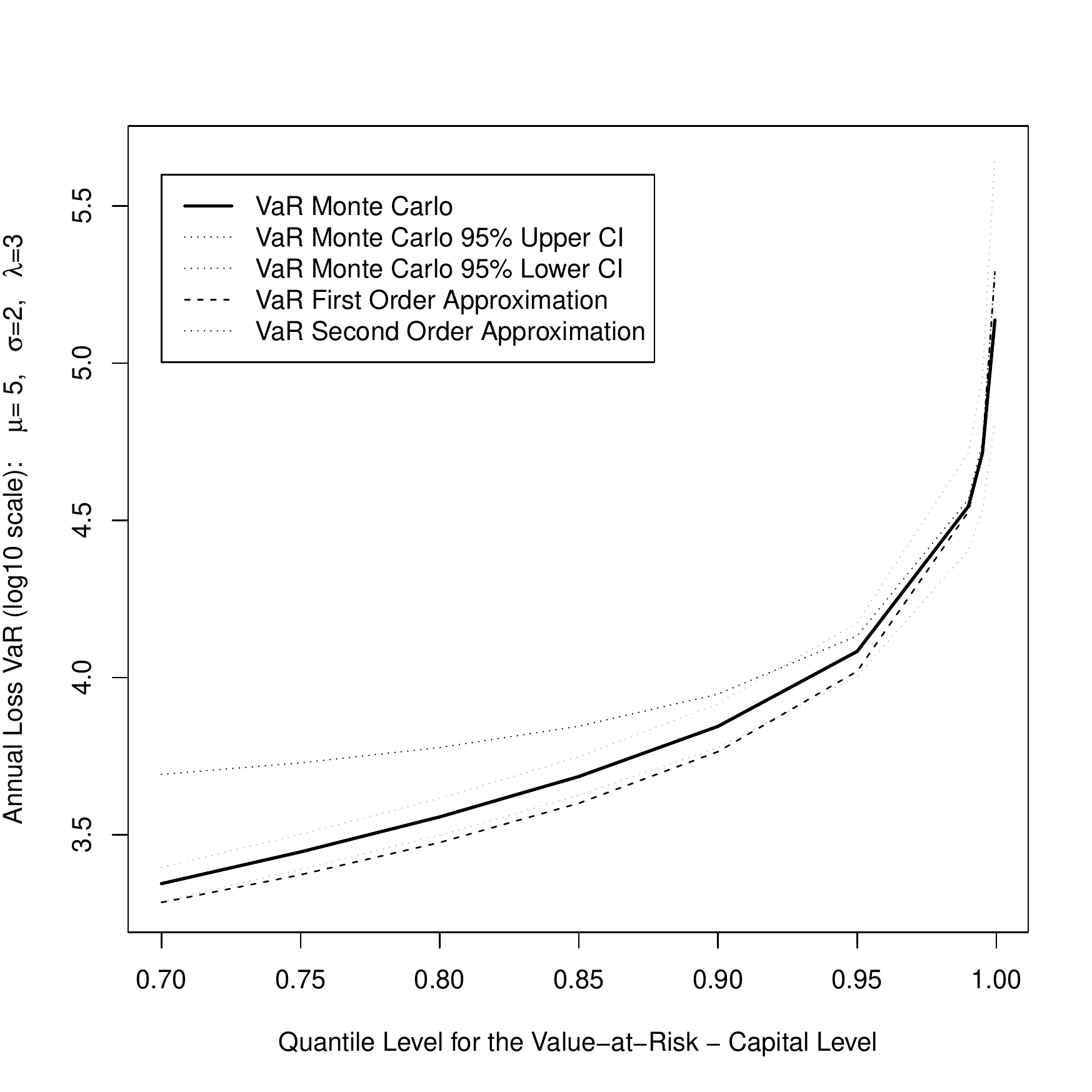}
	\end{minipage}%
	\caption{VaR approximation for the Poisson-Log-Normal example.}%
	\label{fig:VaRLN}%
\end{figure}  
%
%\begin{figure}[h]
%\label{RHIE_Chpt:LN}
%\includegraphics[height=8cm, width=\textwidth] {Figures/LDA_MCvsSLA_PoissonLogNormal.pdf}
%\caption{Annual Loss VaR Capital Estimate versus quantile level for Poisson-Log Normal LDA Risk Process.
%\textbf{Top Plot:} Severity distribution $\mu = 1, \sigma = 0.5, \lambda = 3$.
%\textbf{Bottom Plot:} Severity distribution $\mu = 1, \sigma = 5, \lambda = 3$.}
%\end{figure}
}
\end{exam}
%%%%%%%%%%%%%%%%%%%%%%%%%%%%%%%%%%%%%%%%%%%%%%%%%%%%%%%%%%%%%%%%%%%%%%%%%%%%%%%%%%%%%%%%%%%%%%%

\begin{figure}%
	\centering
	\begin{minipage}{8cm}%
		\includegraphics[width=\textwidth, height=8cm]{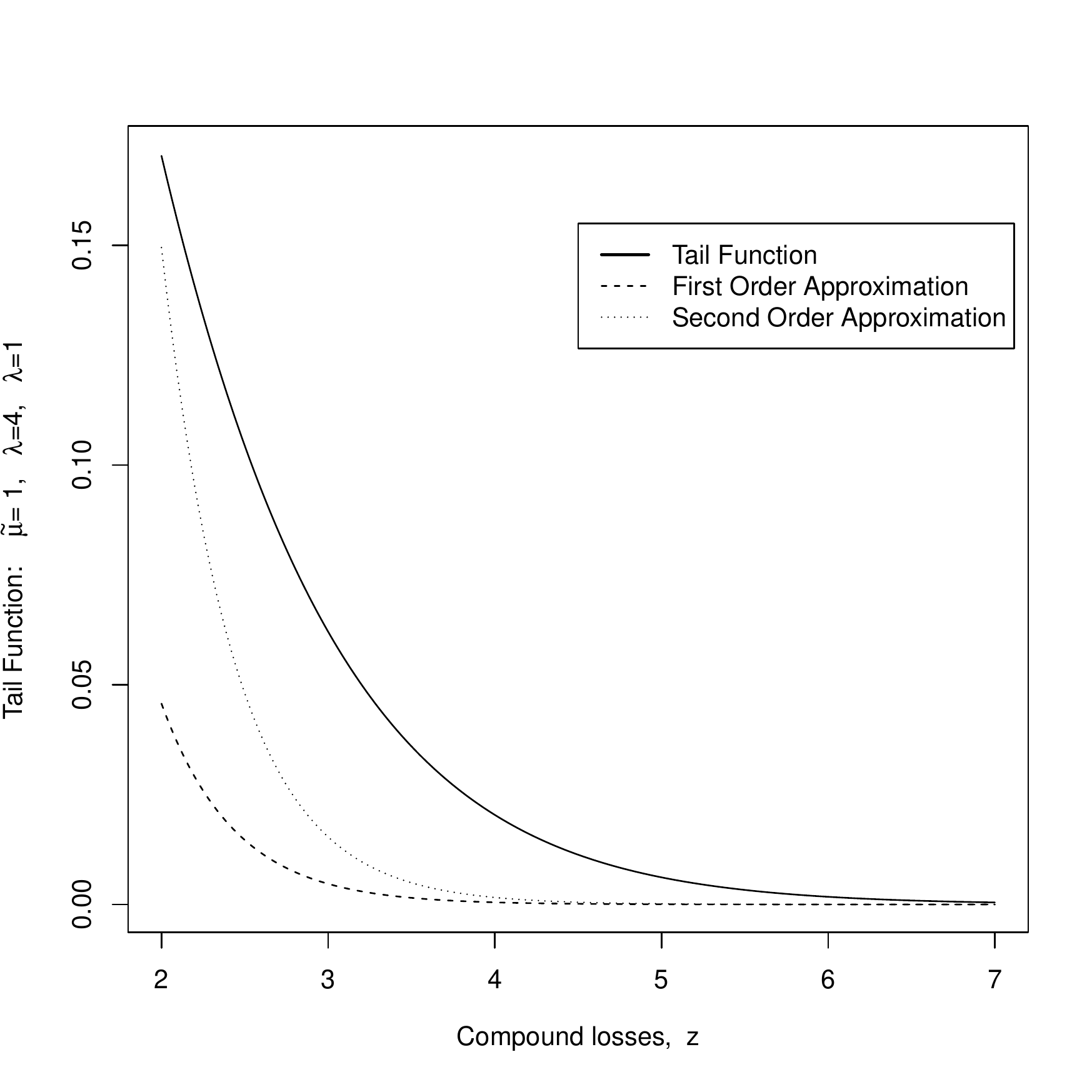}
	\end{minipage}%
	\qquad
	\begin{minipage}{8cm}%
		\includegraphics[width=\textwidth, height=8cm]{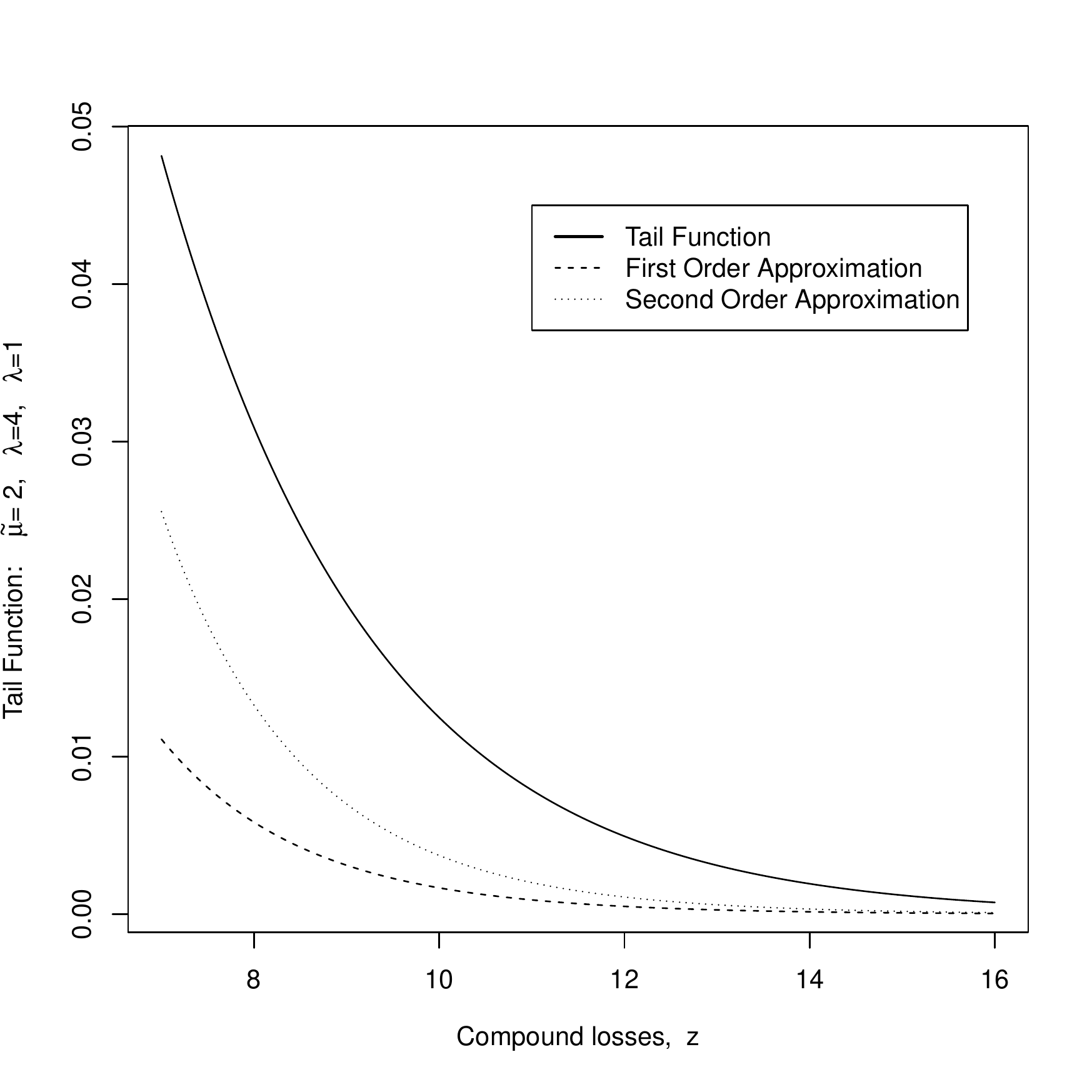}
	\end{minipage}%
	\qquad
	\begin{minipage}{8cm}%
		\includegraphics[width=\textwidth, height=8cm]{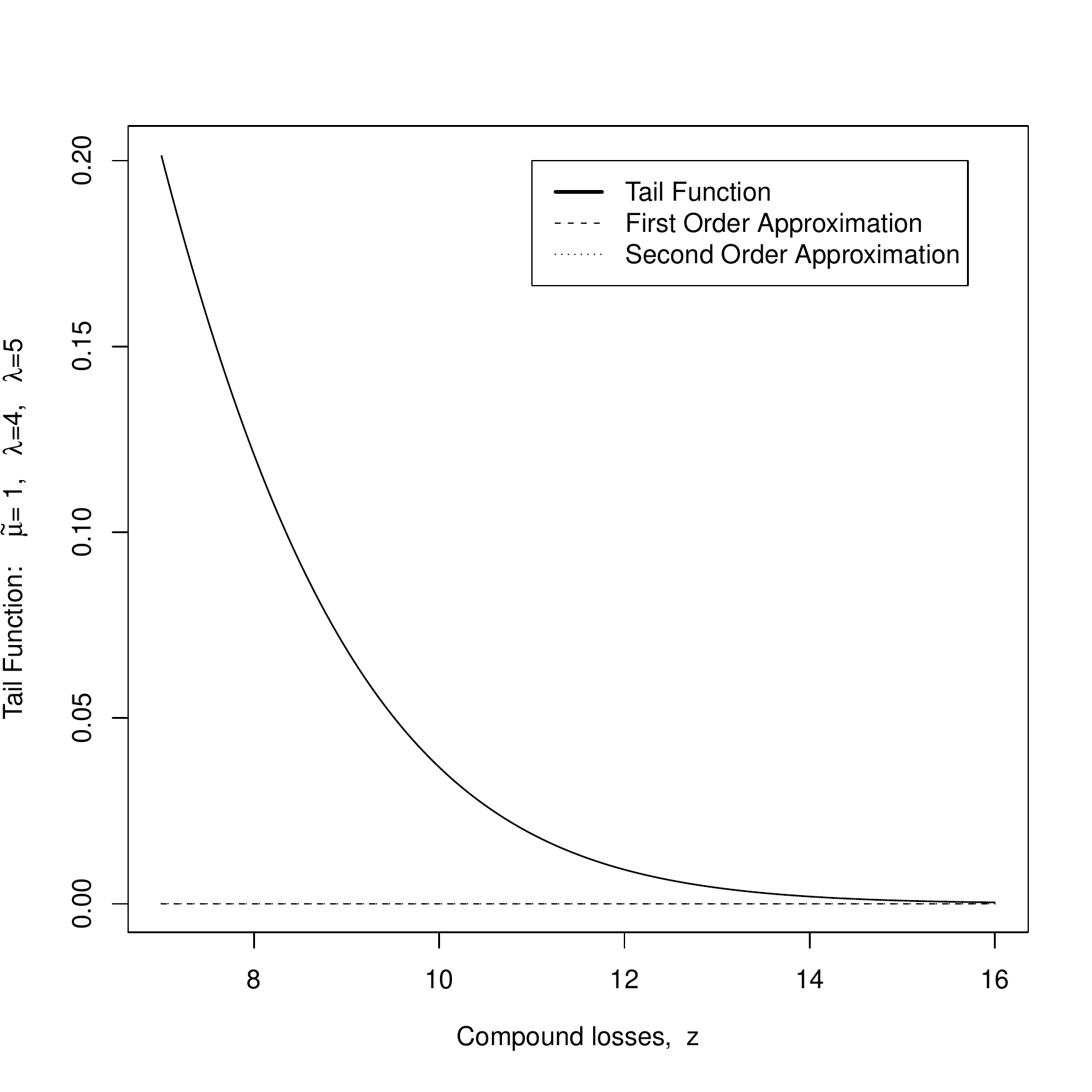}
	\end{minipage}%
	\qquad
	\begin{minipage}{8cm}%
		\includegraphics[width=\textwidth, height=8cm]{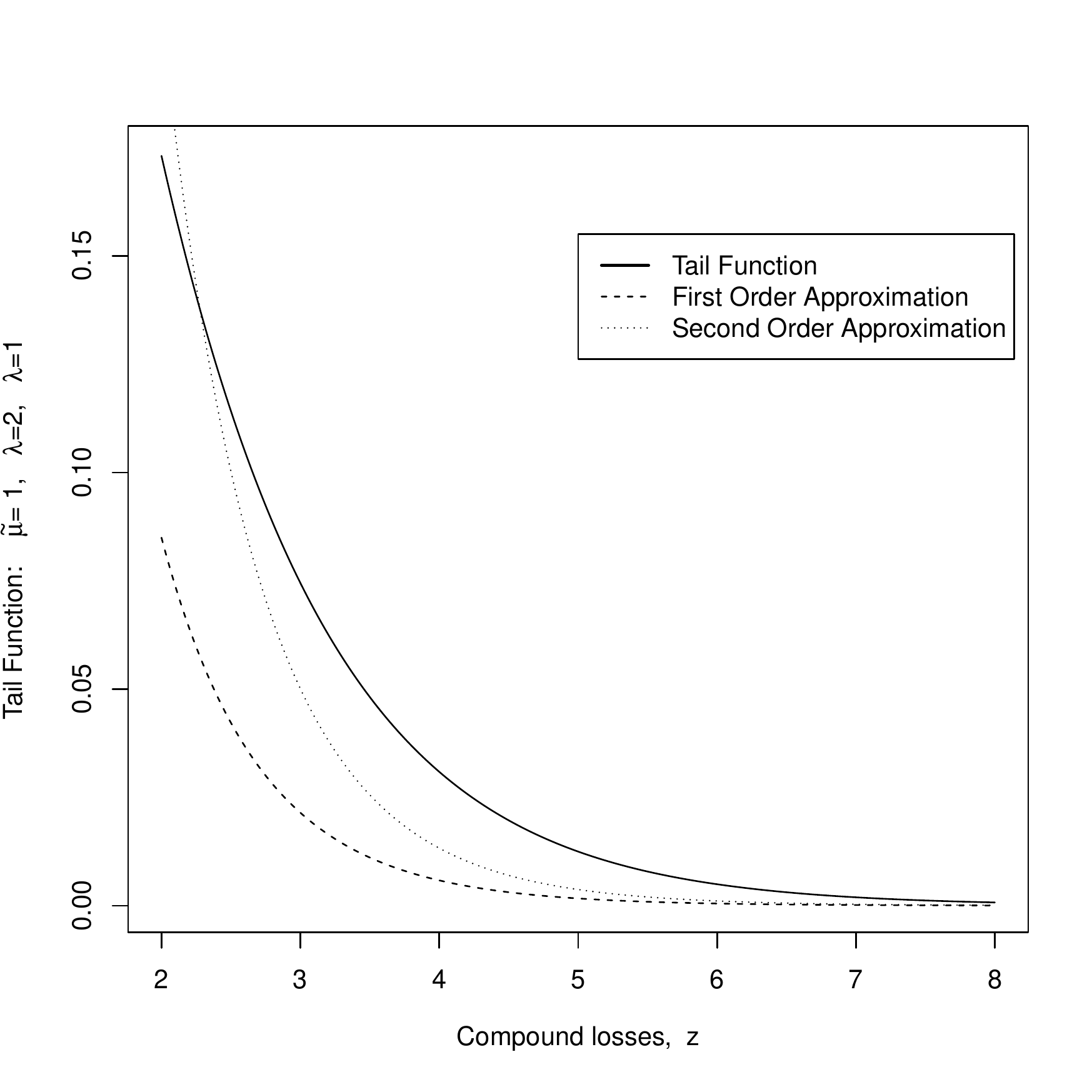}
	\end{minipage}%
	\caption{Tail function approximation for the Poisson-Inverse-Gaussian example.}%
	\label{fig:TailFunctionIG}%
\end{figure}

\begin{figure}%
	\centering
	\begin{minipage}{8cm}%
		\includegraphics[width=\textwidth, height=8cm]{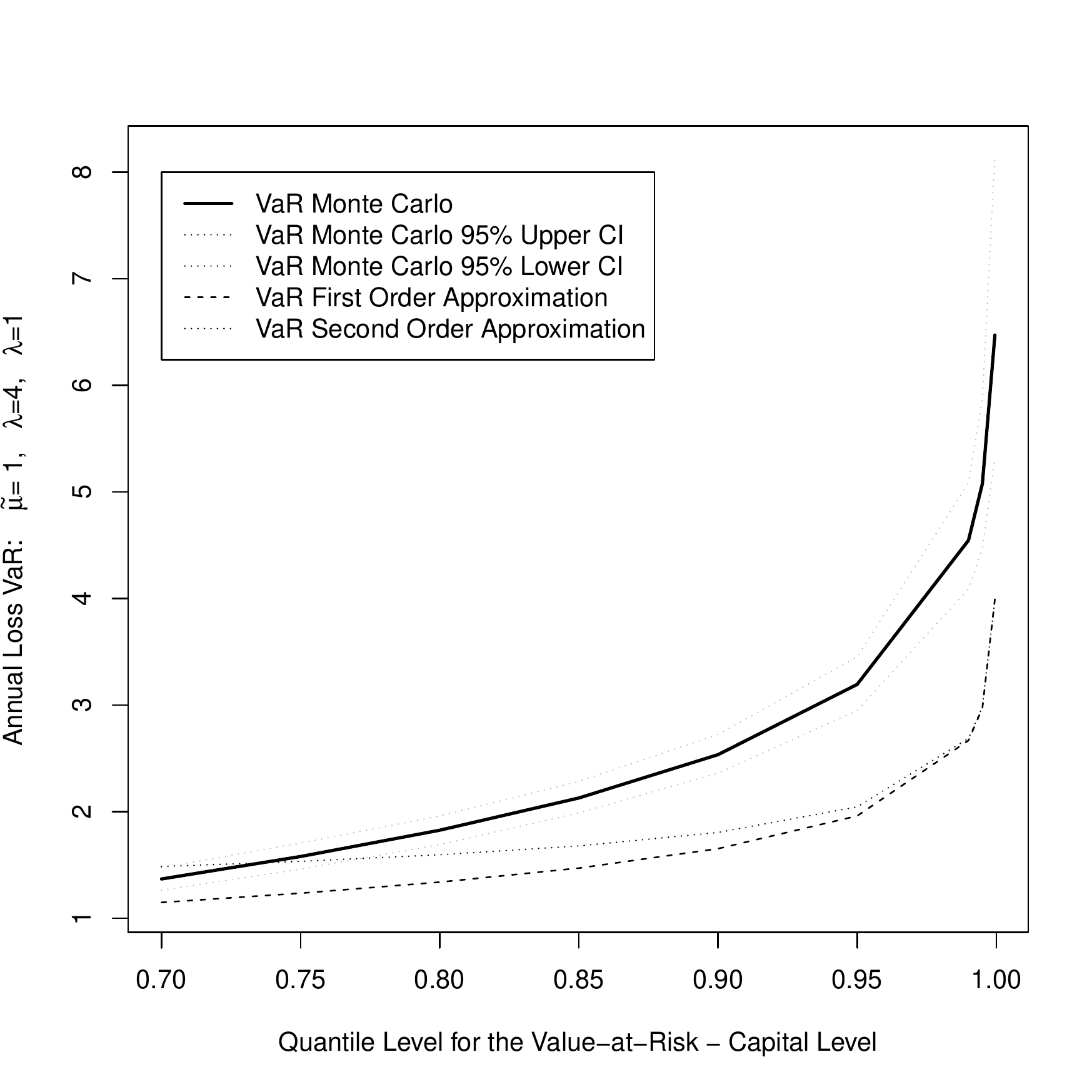}
	\end{minipage}%
	\qquad
	\begin{minipage}{8cm}%
		\includegraphics[width=\textwidth, height=8cm]{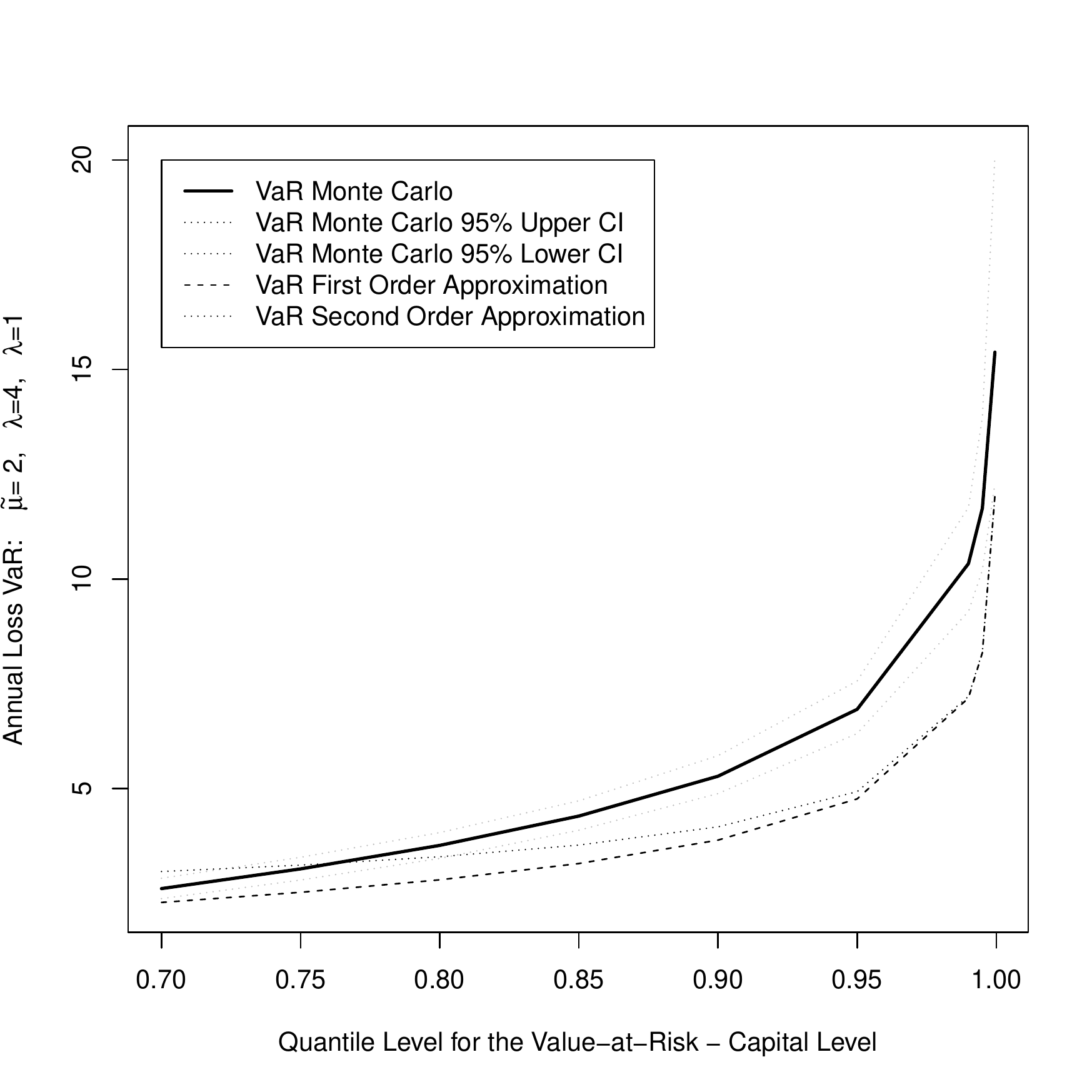}
	\end{minipage}%
	\qquad
	\begin{minipage}{8cm}%
		\includegraphics[width=\textwidth, height=8cm]{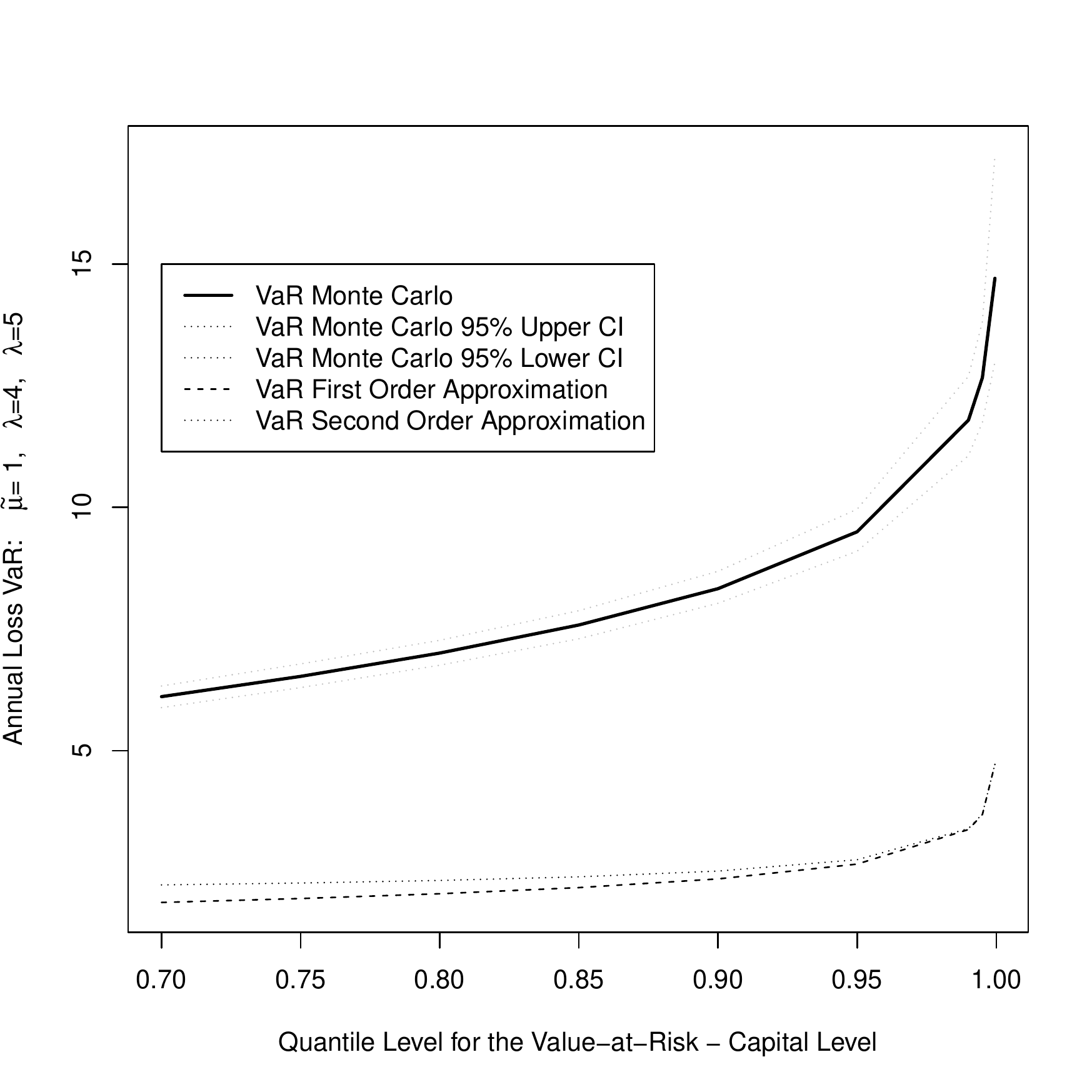}
	\end{minipage}%
	\qquad
	\begin{minipage}{8cm}%
		\includegraphics[width=\textwidth, height=8cm]{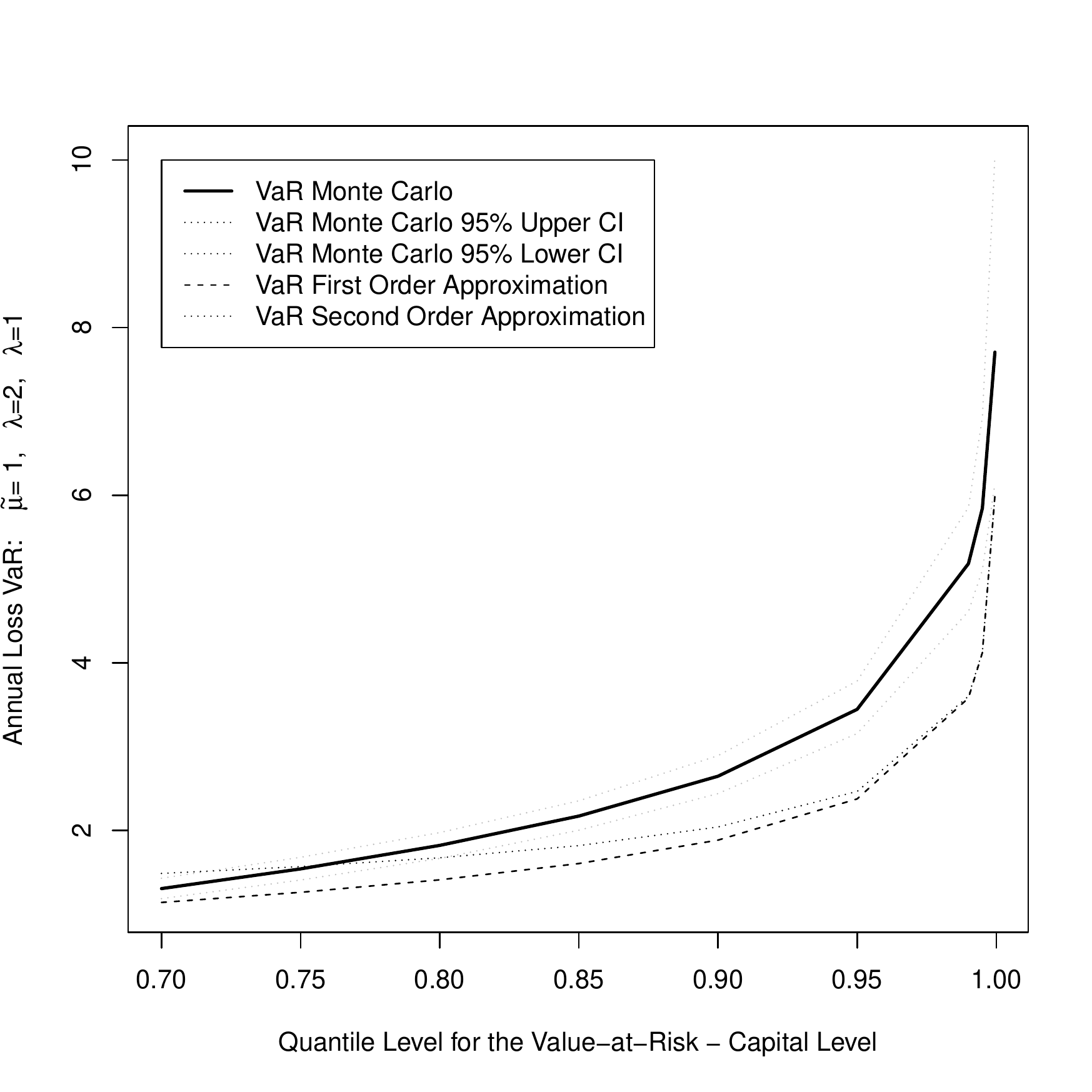}
	\end{minipage}%
	\caption{VaR approximation for the Poisson-Inverse-Gaussian example.}%
	\label{fig:VaRIG}%
\end{figure}

\begin{figure}%
	\centering
	\begin{minipage}{8cm}%
		\includegraphics[width=\textwidth, height=8cm]{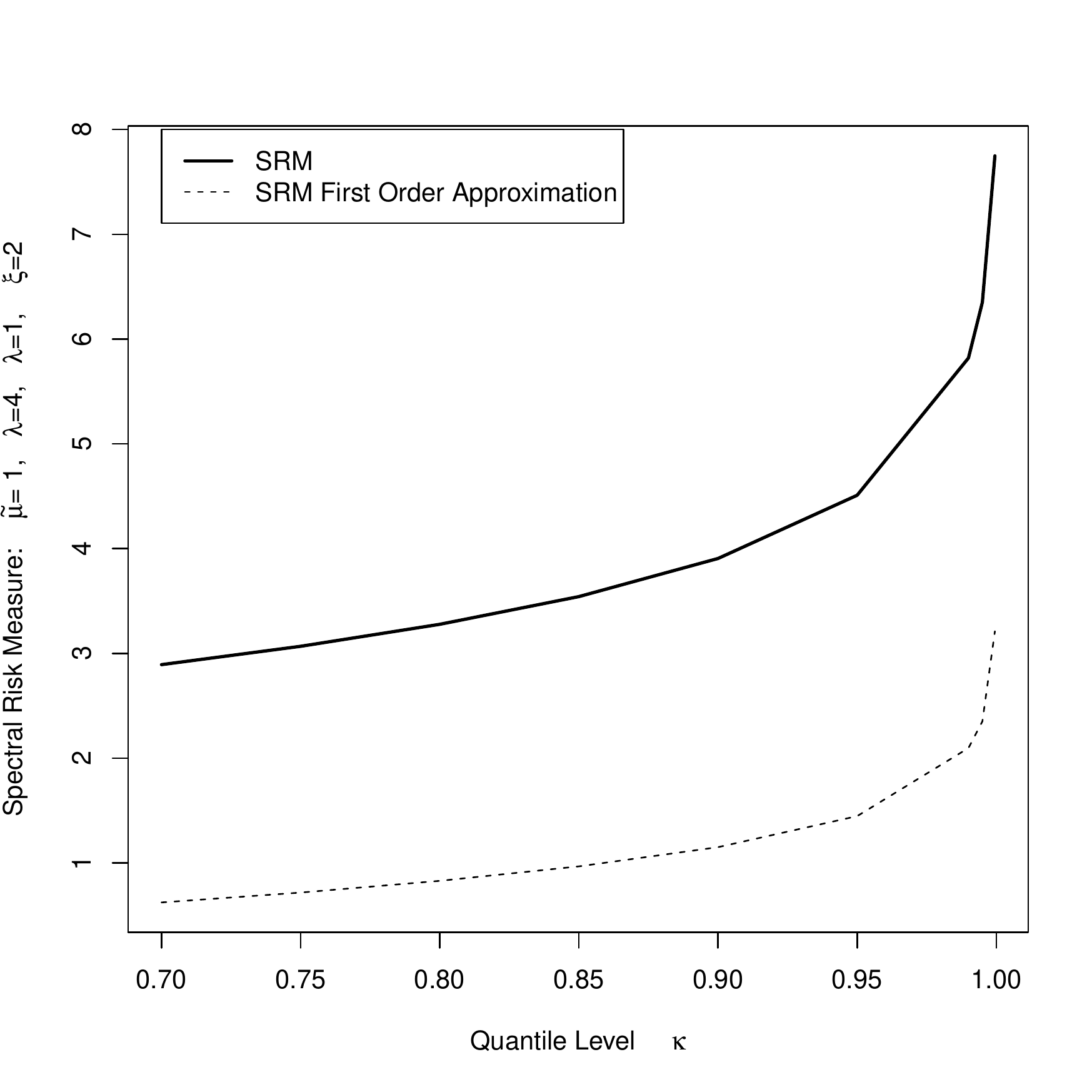}
	\end{minipage}%
	\qquad
	\begin{minipage}{8cm}%
		\includegraphics[width=\textwidth, height=8cm]{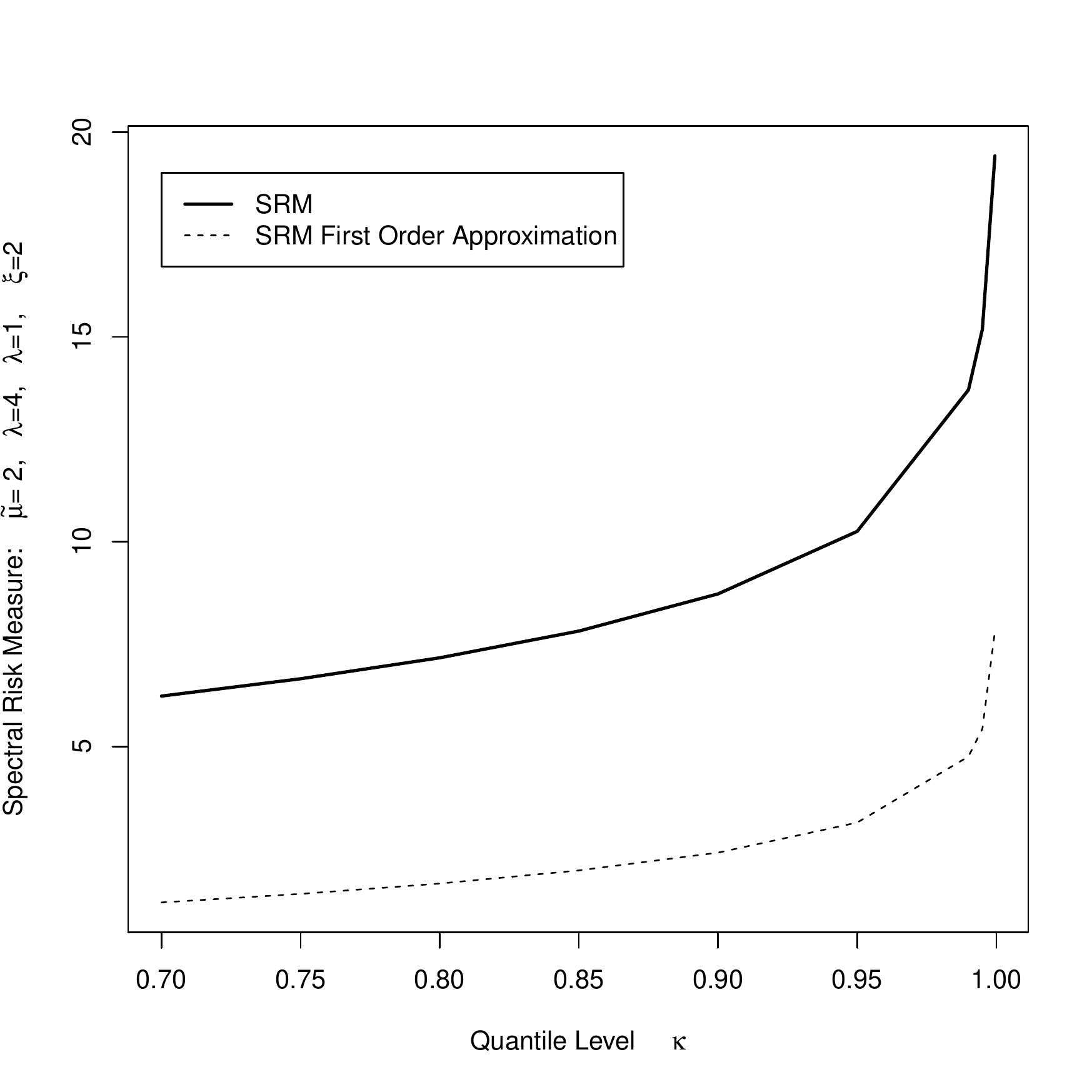}
	\end{minipage}%
	\qquad
	\begin{minipage}{8cm}%
		\includegraphics[width=\textwidth, height=8cm]{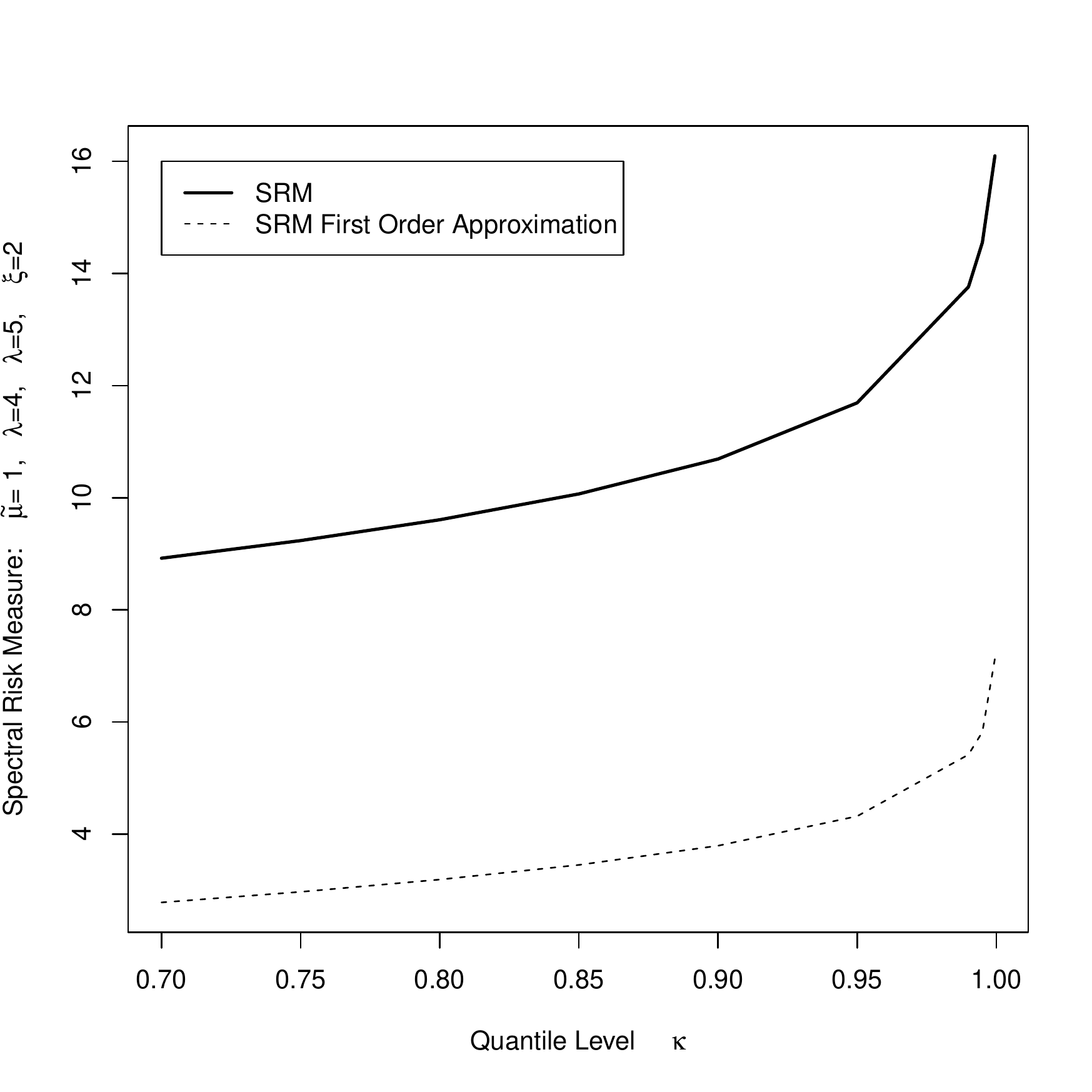}
	\end{minipage}%
	\qquad
	\begin{minipage}{8cm}%
		\includegraphics[width=\textwidth, height=8cm]{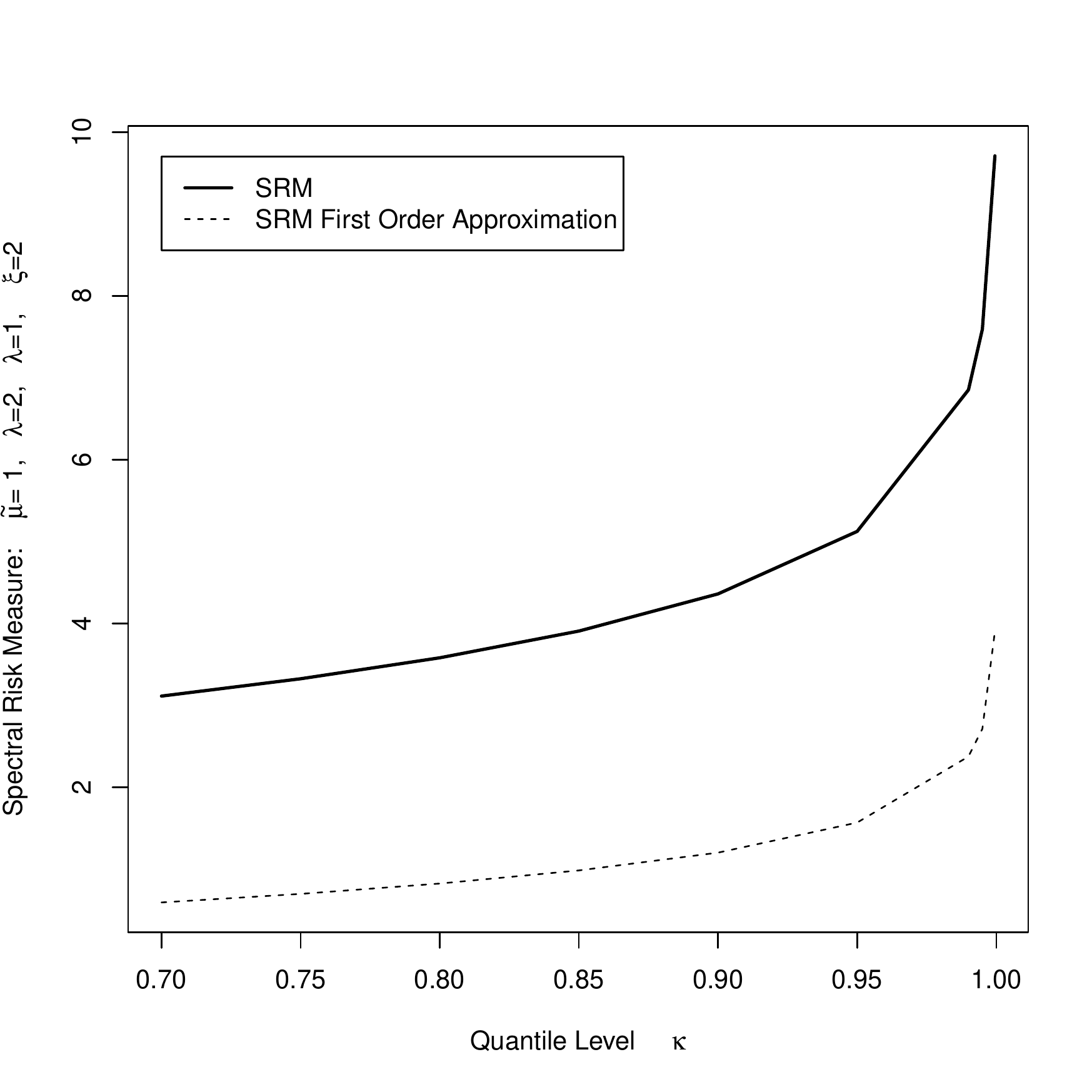}
	\end{minipage}%
	\caption{SRM approximation for the Poisson-Inverse-Gaussian example.}%
	\label{fig:SRMIG}%
\end{figure}

\FloatBarrier
%%%%%%%%%%%%%%%%%%%%%%%%%%%%%%%%%%%%%%%%%%%%%%%%%%%%%%%%%%%%%%%%%%%%%%%%%%%%%%%%%%%%%%%%%%%%%%%
\begin{exam}[Single Risk LDA Poisson-Inverse-Gaussian Family]{
Consider the heavy tailed severity model, selected to model the sequence of i.i.d. losses in each year $t$, denoted $\left\{X_i(t)\right\}_{i=1:N_t}$, and chosen to be an Inverse-Gaussian distribution $X_i \sim IGauss(\widetilde{\mu}, \widetilde{\lambda})$ where the two parameters in this model correspond to parameterizing the scale of the distribution for the severity $\widetilde{\mu}$ and the scale of the distribution $\widetilde{\lambda}$. The survival and quantile functions of the severity are given by
\begin{align*}
f(x;\widetilde{\mu},\widetilde{\lambda}) &= \left(\frac{\widetilde{\lambda}}{2\pi} \right)^{1/2} x^{-3/2} \exp \left\{ -\frac{\widetilde{\lambda} (x - \widetilde{\mu})^2}{2 \widetilde{\mu}^2x} \right\} \\
F(x;\widetilde{\mu},\widetilde{\lambda}) &= \Phi\left(\sqrt{\frac{\widetilde{\lambda}}{x}} \left( \frac{x}{\widetilde{\mu}}-1\right) \right) + \exp\left( \frac{2\widetilde{\lambda}}{\widetilde{\mu}}\right) \Phi\left(-\sqrt{\frac{\widetilde{\lambda}}{x}} \left( \frac{x}{\widetilde{\mu}}+1\right) \right).
\end{align*}

Therefore the closed form SLA for the VaR risk measure at level $\alpha$ would be presented in this case under first and second order approximations for the annual loss $Z = \sum_{n=1}^N X_i$ according to Equations (\ref{Eqn:VaRLNSLA1_2}) and (\ref{Eqn:VaRLNSLA2_2}), respectively
\begin{align} 
	\mathrm{VaR}_{\alpha}\left[Z\right] &= F^{-1}\left(\frac{1-\alpha}{\lambda}\right) \label{Eqn:VaRLNSLA1_2} \\
	\mathrm{VaR}_{\alpha}\left[Z\right] &= F^{-1}\left( 1- \frac{1- \alpha}{\lambda}\{ 1+\widetilde{c}_\beta g_1(F^{-1}(\widetilde{\alpha}))  \}^{-1}\right) \label{Eqn:VaRLNSLA2_2}
\end{align}
where $\widetilde{\alpha} = 1- (1-\alpha)/\lambda $, $g_1(x) = \frac{f(x)}{1-F(x)}$ and $\widetilde{c}_\beta =\widetilde{\mu}\lambda$.

Since the Inverse-Gaussian family is closed under convolution, ie,
$$X_i \sim IGauss(\widetilde{\mu}, \widetilde{\lambda}) \Rightarrow S_n = \sum_{i=1}^n X_i \sim IGauss(n \widetilde{\mu}, n^2\widetilde{\lambda}),$$
we can calculate the distribution of the compound process analiticaly (see the comparison with the approximations on Figure \ref{fig:TailFunctionIG}). The drawback of this family is that there is no closed form for the inverse cdf, which obliges us to resort to a numerical procedure for obtaining the quantiles, fortunately this is efficient and accurate for this class of models. For this model we also present first and second order approximations for the VaR and SRM for different choices of parameters on Figures \ref{fig:VaRIG}, \ref{fig:SRMIG}.
}
\end{exam}
%%%%%%%%%%%%%%%%%%%%%%%%%%%%%%%%%%%%%%%%%%%%%%%%%%%%%%%%%%%%%%%%%%%%%%%%%%%%%%%%%%%%%%%%%%%%%%%

One can see that even in these relatively simple examples, depending on the values of the parameters in the LDA risk model, the asymptotic VaR approximation may or may not be accurate at quantile levels of interest to risk management. Therefore, even small amounts of parameter uncertainty in the LDA model estimation may manifest in significanlty different accuracies in the SLA capital estimates. 

In the Poisson-Log-Normal Example, we can see on Figure \ref{fig:VaRLN} that the volatility of the severities, $\sigma$, play a very important role on the accuracy of the approximations. It is also important to note that although in all the cases the second order approximations does not perform very well for quantiles bellow the 95-th percentile above this threshold it usually performs better than the first order.

For the Poisson-Inverse-Gaussian case the greatest sensitivity is clearly on the rate $\lambda$ (see, for example, the bottom-left plot on Figure \ref{fig:VaRIG}). In difference to the Log-Normal example, the second order approximation seems to perform always better than the first order, but none of them are sufficiently close to any of the ``true" (Monte Carlo) VaR. 

These results therefore serve to illustrate the importance of understanding and developing further studies on the rate of convergence of these asymptotic single loss approximations. This will help to guide in the understanding of when they can be reliably utilised in practice.

\FloatBarrier

%In the top subplot in Figure \ref{RHIE_Chpt:LN} it is evident that the accuracy of the capital estimate formed from the SLA asymptotic result will be poor for this range of parameter values in the LDA model, where as the accuracy from the second subplot with the different risk profile parameters is accurate for quantiles beyond the 80-th percentile. 
%
%However, in general since the rate of convergence is still an active topic of research for such approximations, the only way to ensure accuracy of such methods for a given set of estimated / specified parameters in practice is to complement these approximations with a numerical solution or comparison. In general such an approach will require more sophisticated numerical procedures for more challenging risk process settings. Due to the very simple Monte Carlo approach we present in this example, we required  5,000,000 samples of annual years so that the Monte Carlo accuracy was sufficient. However, we note that as $\lambda$ decreases or as the tail heavyness of the severity model increases, this number of samples will need to significantly increase under a simple Monte Carlo method for the same level or accuracy in the estimation of the tail functionals. Therefore, we provide below a range of alternative Monte Carlo strategies to reduce the required computational cost whilst preserving accuracy of the estimation.

%%%%%%%%%%%%%%%%%%%%%%%%%%%%%%%%%%%%%%%%%%%%%%%%%%%%%%%%%%%%%%%%%%%%%%%%%%%%%%%%%%%%%%%%%%%%%%%
%%%%%%%%%%%%%%%%%%%%%%%%%%%%%%%%%%%%%%%%%%%%%%%%%%%%%%%%%%%%%%%%%%%%%%%%%%%%%%%%%%%%%%%%%%%%%%%
\section{Conclusions}
This work presented an extensive journey through some of the fundamental concepts necessary to the understanding of approximation of OpRisk measures under the Loss Distributional Approach (LDA). We discussed some of the most important classes of severities distributions classified by their right tail properties (such as regularly varying and long tailed) and their interrelations, which we think can help practitioners in the choice of the apropriate heavy tailed model.
Due to the complex nature of these models, assymptotic results are often required for the calculation of risk measures (VaR and Expected Shortfall, for example) and a review of how these so-called First and Second-Order approximations are obtained was provided alogn with a summary of several key results. Finally, for two popular LDA models First and Second-Order approximations can give precise results, but are highly sensitive to the parameters.

%%%%%%%%%%%%%%%%%%%%%%%%%%%%%%%%%%%%%%%%%%%%%%%%%%%%%%%%%%%%%%%%%%%%%%%%%%%%%%%%%%%%%%%%%%%%%%%
%%%%%%%%%%%%%%%%%%%%%%%%%%%%%%%%%%%%%%%%%%%%%%%%%%%%%%%%%%%%%%%%%%%%%%%%%%%%%%%%%%%%%%%%%%%%%%%

%%%%%%%%%%%%%%%%%%%%%%%%%%%%%%%%%%%%%%%%%%%%%%%%%%%%%%%%%%%%%%%%%%%%%%%%%%%%%%%%%%%%%%%%%%%%%%%
%%%%%%%%%%%%%%%%%%%%%%%%%%%%%%%%%%%%%%%%%%%%%%%%%%%%%%%%%%%%%%%%%%%%%%%%%%%%%%%%%%%%%%%%%%%%%%%
\bibliographystyle{ieeetr}
\bibliography{SumyConferenceRefs}

\end{document}